\documentclass[manuscript]{emulateapj}
\usepackage{amsmath}
\usepackage{amssymb}
\usepackage{natbib}
\usepackage{hyperref}
\usepackage{breakurl}
\usepackage{xcolor}
\usepackage{enumitem}
\setlist{leftmargin=5.5mm}
\usepackage{lineno}

\newcommand{\mail}{lilirayhk@gmail.com}

\newcommand{\lum}{\,erg\,s$^{-1}$}

\newcommand{\cm}{\,cm$^{-2}$}
\newcommand{\nh}{$N_\mathrm{H}$}

\newcommand{\src}{3FGL~J0427.9$-$6704}
\shorttitle{Flare-dominated accretion mode of \src}
\shortauthors{Li et al.}

\begin{document}
\title{The flare-dominated accretion mode of a radio-bright candidate transitional millisecond pulsar}
\author{
Kwan-Lok Li\altaffilmark{1,2}, 
Jay Strader\altaffilmark{1}, 
James C. A. Miller-Jones\altaffilmark{3}, 
Craig O. Heinke\altaffilmark{4}, 
Laura Chomiuk\altaffilmark{1}
}

\altaffiltext{1}{Center for Data Intensive and Time Domain Astronomy, Department of Physics and Astronomy, Michigan State University, East Lansing, MI 48824, USA; \href{mailto:\mail}{\mail} (KLL)}
\altaffiltext{2}{Institute of Astronomy, National Tsing Hua University, Hsinchu 30013, Taiwan}
\altaffiltext{3}{International Centre for Radio Astronomy Research, Curtin University, GPO Box U1987, Perth, WA 6845, Australia}
\altaffiltext{4}{Department of Physics, University of Alberta, CCIS 4-183, Edmonton, AB T6G 2E1, Canada}

\begin{abstract}

We report new simultaneous X-ray and radio continuum observations of \src, a candidate member of the enigmatic class of transitional millisecond pulsars. These XMM-Newton and Australia Telescope Compact Array observations of this nearly edge-on, eclipsing low-mass X-ray binary were taken in the sub-luminous disk state at an X-ray luminosity of $\sim10^{33}\,(d/2.3\,{\rm kpc})^2\,$\lum. Unlike the few well-studied transitional millisecond pulsars, which spend most of their disk state in a characteristic \emph{high} or \emph{low} accretion mode with occasional flares, \src\ stayed in the \emph{flare} mode for the entire X-ray observation of $\sim 20$~hours, with the brightest flares reaching $\sim 2 \times 10^{34}$\lum. The source continuously exhibited flaring activity on time-scales of $\sim 10$--100~sec in both the X-ray and optical/UV. No measurable time delay between the X-ray and optical/UV flares is observed, but the optical/UV flares last longer, and the relative amplitudes of the X-ray and optical/UV flares show a large scatter. The X-ray spectrum can be well-fit with a partially-absorbed power-law ($\Gamma \sim 1.4$--1.5), perhaps due to the edge-on viewing angle.  Modestly variable radio continuum emission is present at all epochs, and is not eclipsed by the secondary, consistent with the presence of a steady radio outflow or jet. The simultaneous radio/X-ray luminosity ratio of \src\ is higher than any known transitional millisecond pulsars and comparable to that of stellar-mass black holes of the same X-ray luminosity, providing additional evidence that some neutron stars can be as radio-loud as black holes.

\end{abstract}

\section{Introduction}
\label{sec:intro}
Transitional millisecond pulsars (tMSPs) are a new sub-class of neutron star low-mass X-ray binaries (NS-LMXBs) that becomes observationally known in the last decade (see \citealt{2018IAUS..337...47J} for the time-line of some of the most significant events of the class). 
Unlike typical accreting millisecond X-Ray pulsars (AMXPs), these systems switch between distinct states of 
being a pulsar and an LMXB on time-scales that range from weeks to $\sim10+$~years {\citep{2002PASP..114.1359B,2005AJ....130..759T,2009ApJ...703.2017W,2009Sci...324.1411A,2014ApJ...781L...3P,2013Natur.501..517P}.
As the only-known bridge between the radio MSPs and LMXBs, they are widely linked to the standard recycling scenario of neutron stars \citep{1982Natur.300..728A,1982CSci...51.1096R}. 

To date, only three tMSPs are known: PSR J1824-2452I in M28 (a.k.a. M28I; \citealt{2013Natur.501..517P}), PSR J1023+0038 \citep{2009Sci...324.1411A,2014ApJ...781L...3P}, and PSR J1227$-$4853 \citep{2015ApJ...800L..12R}. 
They are all identified as ``redback'' eclipsing millisecond pulsar binaries, in which the MSP is ablating the low-mass companion (median mass of $0.36\,M_\sun$; \citealt{2019ApJ...872...42S}) in a compact orbit (orbital periods of $\lessapprox$ 1 day). 

M28I is currently the only known tMSP that showed a typical X-ray outburst as AMXPs (i.e., $L_x\gtrsim10^{36}$\lum).
In PSRs J1023+0038 and J1227$-$4853, the accretion state is about two orders of magnitude lower ($L_x\lesssim10^{34}$\lum).
In this so-called sub-luminous disk state, at least three accretion modes, namely the \textit{low} (a few $\times 10^{32}$\lum), \textit{high} (a few $\times 10^{33}$\lum), and \textit{flare} modes ($\sim10^{34}$\lum) are observed \citep{2013A&A...550A..89D,2015ApJ...806..148B}. At least one candidate tMSP, 3FGL~J1544.6$-$1125, has been identified via its display of similar accretion modes and its other optical properties \citep{2015ApJ...803L..27B,2017ApJ...849...21B}. Like 3FGL~J1544.6$-$1125, PSR~J1023+0038 and PSR J1227$-$4853 have been observed to emit GeV $\gamma$-rays in the sub-luminous disk states \citep{2014ApJ...790...39S,2015ApJ...806...91J}.
Interestingly, PSR~J1023+0038 also exhibited optical pulsations during the sub-luminous disk state, which makes it the first millisecond pulsar ever detected in optical \citep{2017NatAs...1..854A}. \cite{2019ApJ...882..104P} argued that the pulsed optical emission originates neither from magnetically channelled accretion nor rotation-powered pulsar magnetosphere, but synchrotron emission from the intrabinary shock between the pulsar wind and the accretion disk. This would imply that the rotation-powered activity of a pulsar persists in the sub-luminous disk state.

PSRs~J1023+0038 and J1227$-$4853 are known to spend most of the time in the high (e.g., about $70$\% for PSR~J1023+0038) and low (about $20$\%) modes during the sub-luminous disk state. The mode can promptly switch from high to low in just $\sim10$~sec, and then switch back equally rapidly after $100$--$1000$~sec in the low mode. 
X-ray pulsations have only been detected in the high mode \citep{2015MNRAS.449L..26P,2015ApJ...807...62A}. PSR~J1023+0038 has also been known to enter an extended flare mode occasionally. These extended flaring episodes can last up to $\approx10$~hours in X-rays \citep{2014ApJ...791...77T,2014ApJ...797..111L} and $\approx14$~hours in optical \citep{2018ApJ...858L..12P}. 
Optical pulsations were also detected when PSR~J1023+0038 was in the flare mode \citep{2019ApJ...882..104P}, suggesting that the flares are happening near the neutron star. 
Recently, \cite{2019A&A...622A.211C} found another tMSP candidate, CXOU J110926.4-650224, to flare in X-rays for up to 4.5 hours. The tMSP identity, however, remains questionable because of the insignificant GeV $\gamma$-ray counterpart \citep{2019arXiv190210045T,2019Galax...7...93H}. 

\begin{figure}
\centering
\includegraphics[width=0.47\textwidth]{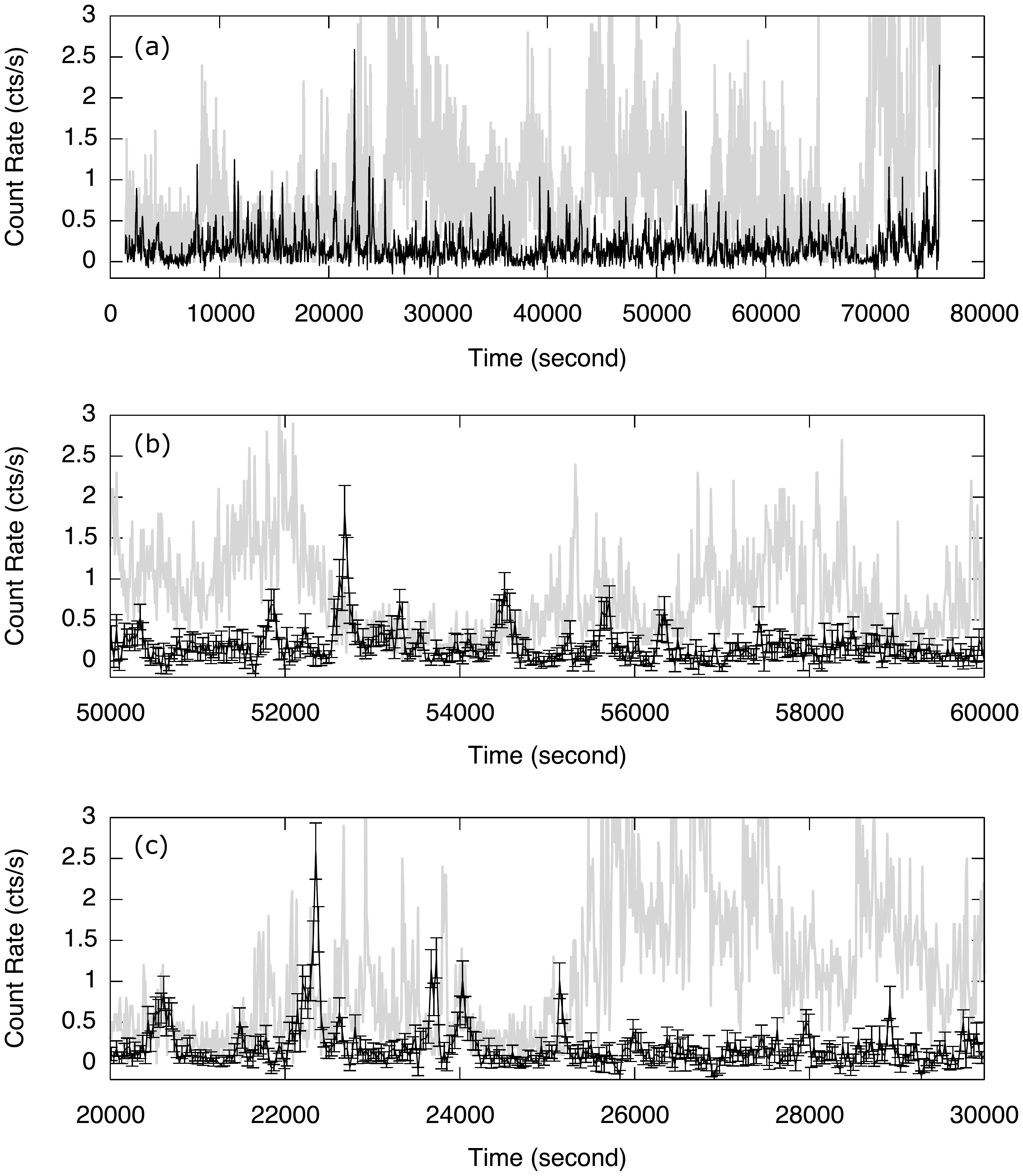}
\caption{(a) The black curve shows the reduced EPIC light curve of \src\ (pn + MOS~1/2), while the gray one indicates the flaring background obtained by pn. (b \& c) Two arbitrary zoomed-in views of (a) with 1-$\sigma$ uncertainties of the EPIC light curve. No correlation is found between the two curves. }
\label{fig:epic_bkg}
\end{figure}

\begin{figure*}[ht]
\centering
\includegraphics[width=\textwidth]{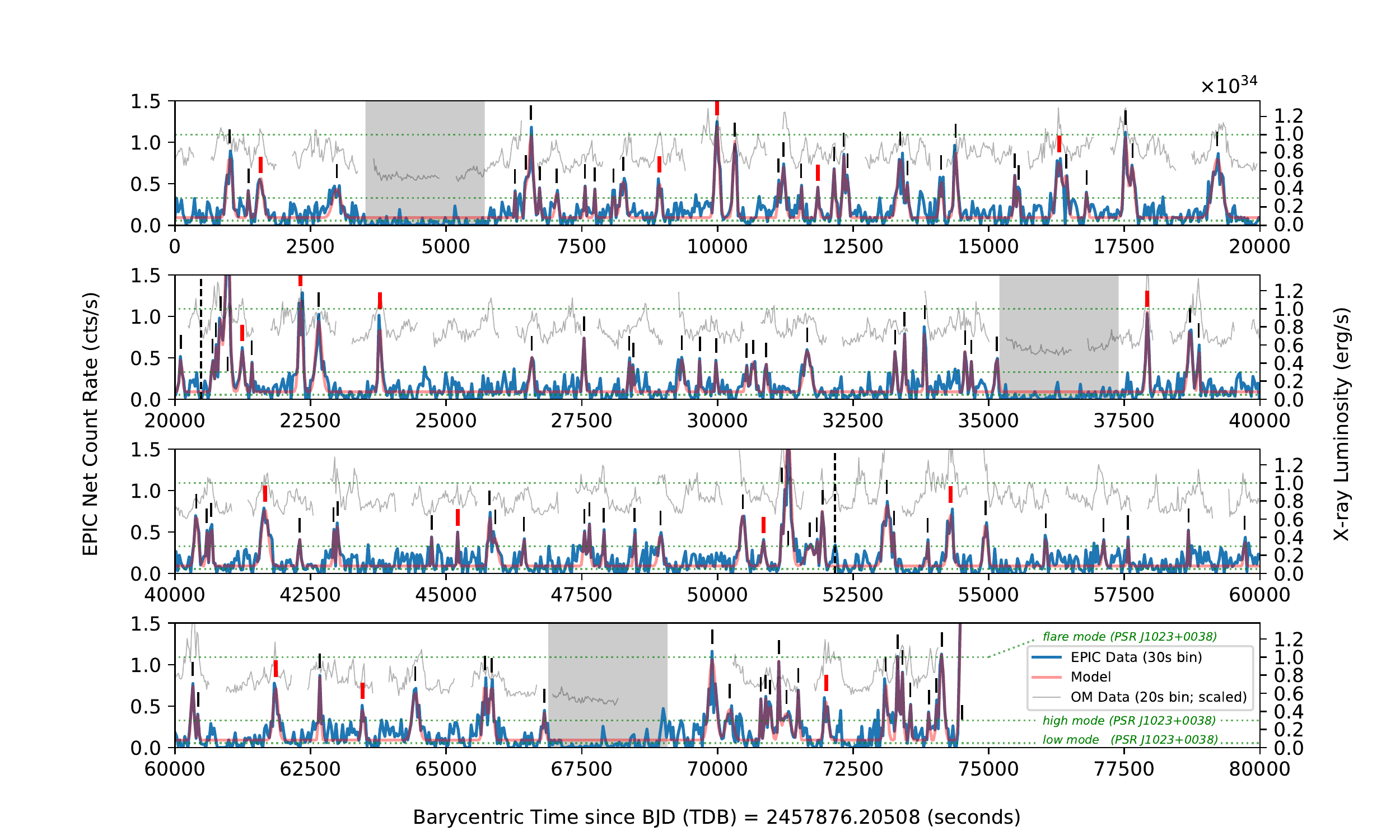}
\caption{The X-ray light curve (0.2--10~keV; blue) of \src\ observed by \textit{XMM-Newton} in May 2017. 
The left y-axis shows the corrected count rate measured by pn and MOS~1/2 (the high flaring background intervals are also included), while the right axis refers to the X-ray luminosity inferred from the best-fit partially-absorbed power-law model in \textit{Average1} (see Table~\ref{tab:xmm_spec} for the best-fit parameters) with $d=2.3$~kpc. 
The three 2200-sec eclipses are all indicated by the gray shaded regions based on the solution in S16 while the two vertical dashed lines indicate the inferior conjunction of the binary. 
The red line is the best-fit Gaussian model for the X-ray flares (see text for details). 
Every flare detected is marked with a vertical bar on the top. Red and thicker bars are used to indicate the 16 representative flares (see Figure \ref{fig:16flare}), and the rest are black in color. 
The gray narrow curve is the OM light curve scaled to fit the figure (see Figure \ref{fig:om_lc} for the original). 
The three horizontal green dashed lines represent the X-ray luminosities (0.3--10~keV) of PSR~J1023+0038 in the flare ($\approx10^{34}$\lum), high ($\approx3\times10^{33}$\lum), and low ($\approx5\times10^{32}$\lum) modes, respectively, which are obtained from \cite{2015ApJ...806..148B}. }
\label{fig:epic_lc}
\end{figure*}

In the radio band, the steep spectrum emission and radio pulsations that were clearly detected in the pulsar state disappear in the sub-luminous disk state \citep{2009Sci...324.1411A,2013Natur.501..517P,2014ApJ...790...39S}. Instead, flat-spectrum radio emission is detected \citep{2011MNRAS.415..235H,2015ApJ...809...13D}. Recently, \cite{2018ApJ...856...54B} found a strong anti-correlation between the radio and X-ray emission in the simultaneous radio and X-ray observations of PSR~J1023+0038---the radio source is bright in the low mode, and faint in the higher mode. These data suggest that a synchrotron-emitting outflow is launched during the low mode. 

The subject of this paper, \src, is a $\gamma$-ray-loud LMXB with an a 8.8-hour orbital period and active accretion, making it very likely to be another sub-luminous tMSP (\citealt{2016ApJ...831...89S}; S16 hereafter). Uniquely among the members of the class, it is a nearly edge-on system with an inclination of $i\approx80^\circ$. Since spectral lines from both the secondary and disk are visible, \src\ is essentially an eclipsing doubled-lined spectroscopy binary, which allowed S16 to measure the masses of both the accreting primary ($M_1\approx1.8 M_\sun$) and secondary ($M_2\approx0.6 M_\sun$).

\src\ was observed in the X-rays (3--79~keV) by \textit{NuSTAR} in 2016. During the 84~kec observations, a non-thermal hard X-ray source was detected, accompanied by three X-ray eclipses. The observation also found strong X-ray variability on time-scales of hundreds of seconds, reminiscent of the mode switching phenomenon seen in tMSPs. In this paper, we report the results of the follow-up \textit{XMM-Newton} and ATCA observations of \src, from which we conclude that \src\ has X-ray and optical/UV flaring properties distinct from that observed in other tMSPs.

\section{Observations and Data Reduction}

\subsection{\textit{XMM-Newton}}
We carried out a 77.5~ksec \textit{XMM-Newton} observation (ObsID: 0801650301; PI: J. Strader), starting from 2017 May 2 at 16:14:17 UT to May 3 at 13:45:57 UT. During the observation, all three {European Photon Imaging Cameras} (EPIC: pn, MOS~1, and MOS~2) were operated under the Full Frame mode (i.e., \textit{PrimeFullWindow} mode) with time resolutions of 73.4~msec and 2.6~sec for the pn and MOS~1/2 cameras, respectively. The {Medium1} filter was used to optimize the X-ray data quality. For the {Optical/UV Monitor Telescope} (OM), {Fast} mode (time resolution of 0.5~sec) with a ``white'' filter (effective wavelength at 406~nm with a 347~nm bandpass width) was used to obtain high-speed optical/UV photometry of the system. We did not use the two {Reflection Grating Spectrometers} owing to the faintness of the source.

\begin{figure*}
\centering
\includegraphics[width=\textwidth]{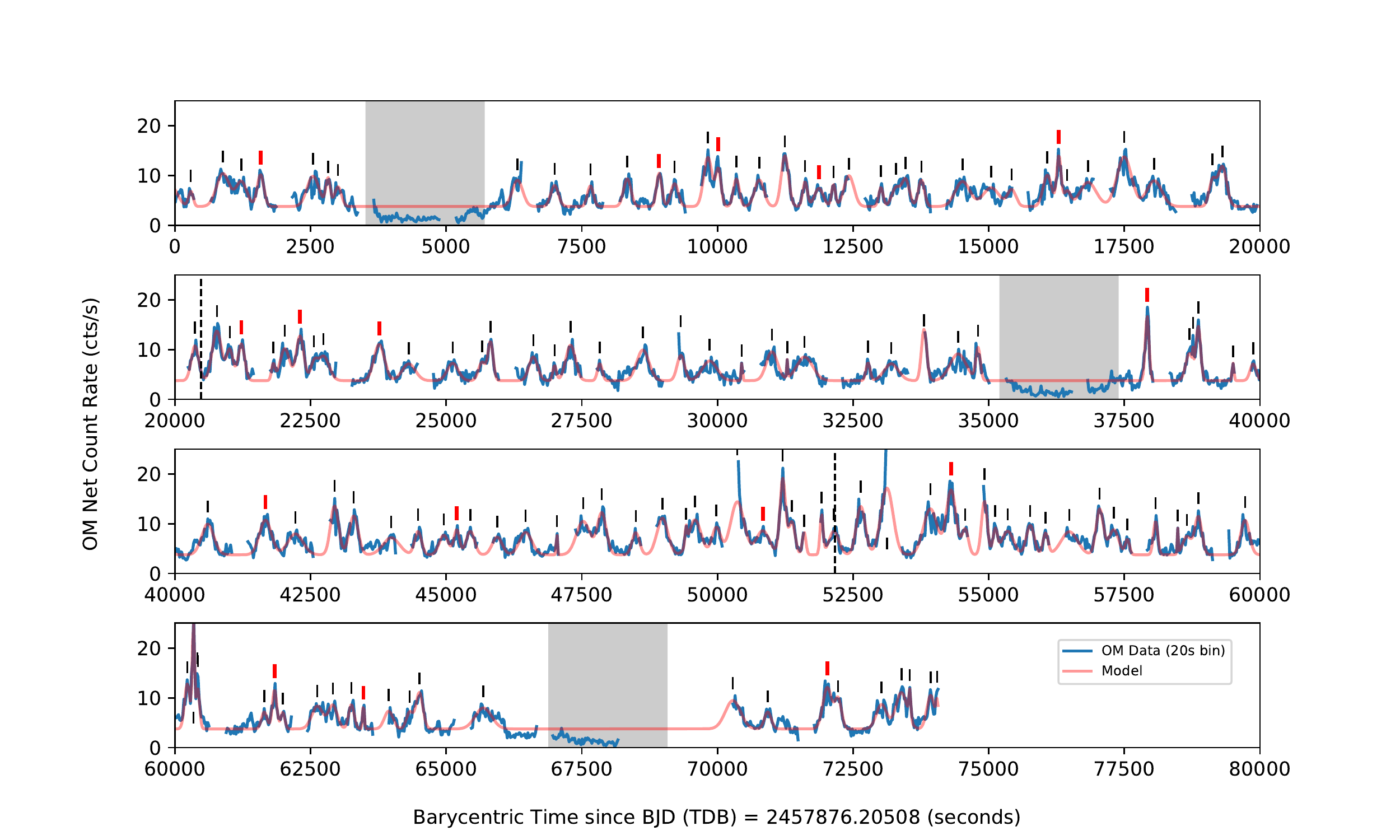}
\caption{The same as in Figure~\ref{fig:epic_lc}, but for the OM optical/UV observations alone. }
\label{fig:om_lc}
\end{figure*}

\label{sec:xmm_re}
We applied the standard analysis tools in the \texttt{Science Analysis System} (SAS; version 15.0.0) and \texttt{HEAsoft} (version 6.22) with the calibration files (CCF) obtained from the on-line \texttt{cifbuild} server\footnote{\url{https://www.cosmos.esa.int/web/xmm-newton/cifbuild}}
to reduce and analyse the \textit{XMM-Newton} data. The SAS task \texttt{xmmextractor} was used for the EPIC's reduction processes, with standard filtering (\texttt{\#XMMEA\_EP \&\& FLAG==0 \&\& PATTERN $\pmb\le$ 4} for pn; and \texttt{\#XMMEA\_EM \&\& FLAG==0 \&\& PATTERN $\pmb\le$ 12} for MOS~1/2). 
To deal with the high flaring particle background, exposure periods with count rates of the pattern zero events higher than 0.4 cts~s$^{-1}$ in 10--12~keV for pn and 0.35 cts~s$^{-1}$ in $>10$~keV for MOS~1/2 (recommended values in the \textit{XMM-Newton} user manual) were removed in the imaging and spectral analyses. 
The flaring background heavily contaminated the pn observation, with an effective exposure time of 24~ksec left after the filtering. Fortunately, the MOS~1/2 observations were less affected, with an effective exposure time 70~ksec for each MOS.
For the extraction regions, circular source regions with optimum radii determined by \texttt{eregionanalyse} were used ($r=26$\arcsec, 24\arcsec, and 27\arcsec\ for pn, MOS~1, and MOS~2, respectively), while source-free annulus regions with inner/outer radii of 60\arcsec/70\arcsec\ were used for the background. 
The full energy band of 0.2--10~keV was used in the analysis, unless mentioned otherwise. 
In Figure \ref{fig:epic_bkg}, we plotted the background-subtracted light curves of \src\ together with the pn flaring background to examine whether the flaring background can contaminate the reduced light curve significantly. It is clearly shown in the figures that the effect is very minor and therefore the whole period of data will be used in our light curve analysis.

The OM images and light curves were reduced and extracted by \texttt{omfchain}. 
There were 48 exposure segments taken in a $22\times23\,{\rm pixels}$ OM window (plate scale of about 0\farcs48 per pixel). Except for the last one with 2280~sec, every segment has an exposure time of 1200 seconds with a $\sim300$-sec observing gap (there is one exception before the last second exposure, which has a gap of $\sim2000$~sec). 
Two sources, the optical counterpart of \src\ and a faint non-variable source (5\farcs5 to the north-east of \src; about 6 times fainter than the target), were detected and well-resolved from each other. 
Aperture photometry with an aperture size of 6 pixels in radius (about 3~FWHM in diameter) was applied to extract the OM light curves of \src. An annulus with inner/outer radii of 7.2 and 15 pixels was used as the background region, in which the back/foreground faint source was excluded with a 6-pixel radius circular region. 
All times are presented in the frame of Barycentric Dynamical Time (TDB), converted by the SAS task \texttt{barycen} with the JPL solar system ephemeris DE200 \citep{1990A&A...233..252S}. 

\begin{figure*}[t]
\centering
\includegraphics[width=\textwidth]{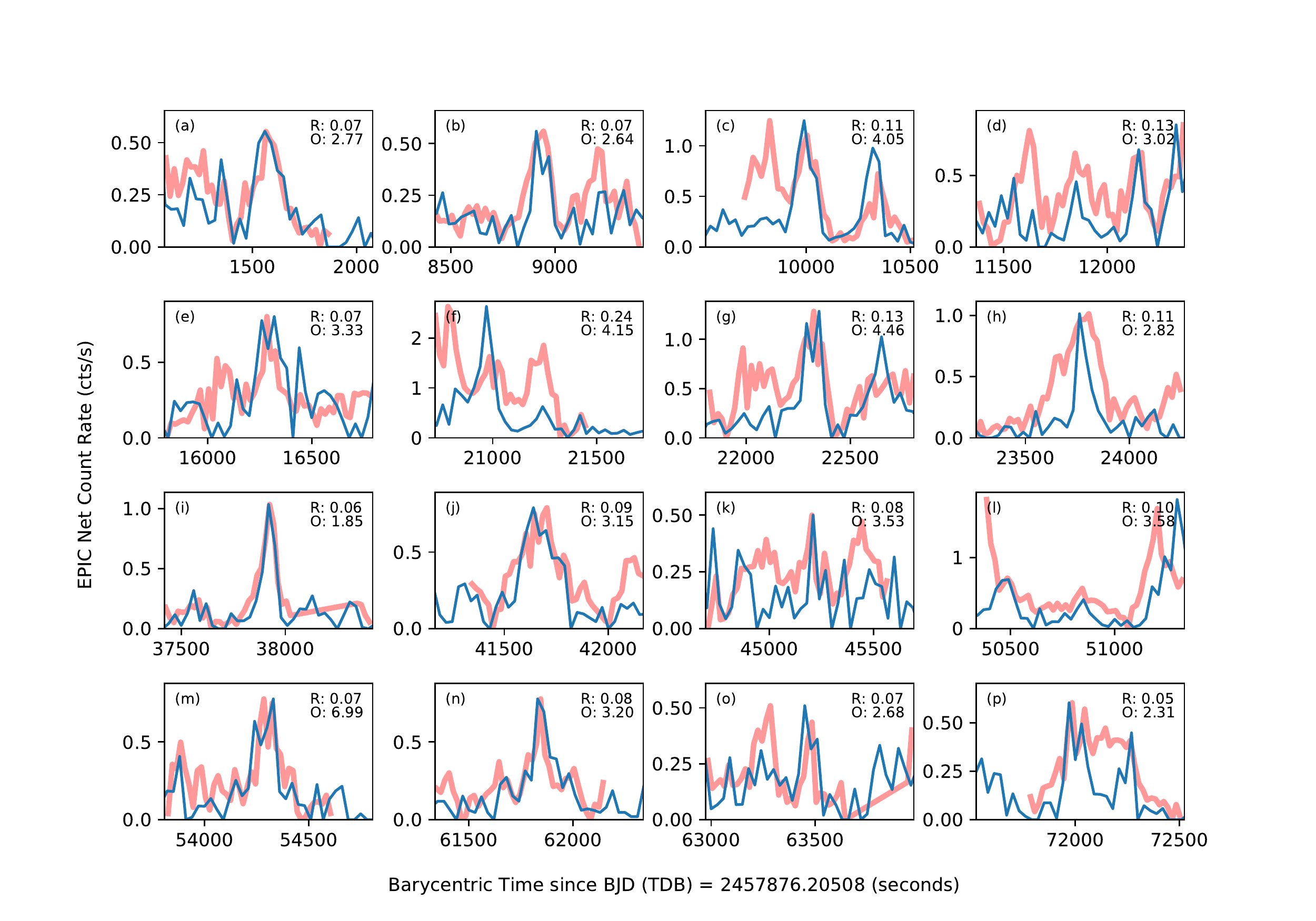}
\caption{The EPIC X-ray (blue and thinner) and optical/UV (red and thicker) light curves of 16 arbitrarily chosen flares to show the range of relationships between the X-ray and optical flares. Each zoom-in view is centred at the peak of the chosen flare. The optical/UV light curves shown are scaled linearly to fit the X-ray ones for easier comparison, i.e., $C_s = R (C_o - O)$, where $C_s$ is the scaled count rate, $C_o$ is the original count rate, $R$ is the X-ray-to-optical/UV flare amplitude ratio, and $O$ is the minimum count rate of the original optica/UV light curve in the interval. $R$ and $O$ can be found in the upper right corner of each plot. }
\label{fig:16flare}
\end{figure*}

\subsection{Australia Telescope Compact Array}

We arranged strictly simultaneous radio observations of \src\ using the Australia Telescope Compact Array (ATCA), observing for as long as the source was above the horizon, from 18:10 UT on 2017 May 2 to 13:07 UT on 2017 May 3, under project codes CX364 and C3170. The array was in the extended 6A configuration, with all six antennas aligned east-west, with a maximum baseline of 6\,km. We used the Compact Array Broadband Backend \citep[CABB;][]{2011MNRAS.416..832W} to observe simultaneously at central frequencies of 5.5 and 9.0\,GHz, with 2048\,MHz of bandwidth in each of the two frequency bands.

We used the standard extragalactic calibrator PKS B1934$-$638 as a bandpass calibrator and to set the flux density scale, and the nearby sources 0355$-$669 and then J0425$-$6646 (after the former calibrator set) to determine the complex gain solutions. We reduced the data using standard procedures within the Common Astronomy Software Application \citep[CASA;][]{2007ASPC..376..127M}. We imaged the calibrated data using Briggs weighting with a robust parameter of 0.5, as a compromise between sensitivity and minimizing sidelobe levels. The source was detected at high significance in both frequency bands.

We imaged several subsets of data, to determine whether the radio emission changed when in eclipse, and how it varied with the changing X-ray count rates defined in Section \ref{sec:spec_analy}. Finally, we made time-resolved light curves. Since the array was in an east-west configuration, the instantaneous {\it uv}-coverage was linear, and not suitable for imaging. We therefore subtracted out all other sources in the field, using our best image made from the entire data set at each frequency, leaving only \src\ in the visibility data, which we fit in the {\it uv}-plane using a point source model fixed at the known source position. We found that ten-minute bins provided the best compromise between time resolution and sensitivity.

\subsubsection{2016 Data}

We previously obtained ATCA radio data of \src\ from 22:29 UT on 2016 Aug 27 to 04:38 UT on 2016 Aug 28, under project code CX365 and in the extended 6C configuration. The receiver setup, calibrators, and imaging were the same as for the 2017 data. A single image was made at each of the central frequencies of 5.5 and 9.0 GHz.
These radio data were not simultaneous with any X-ray observations.

\begin{figure*}
\centering
\includegraphics[width=0.85\textwidth]{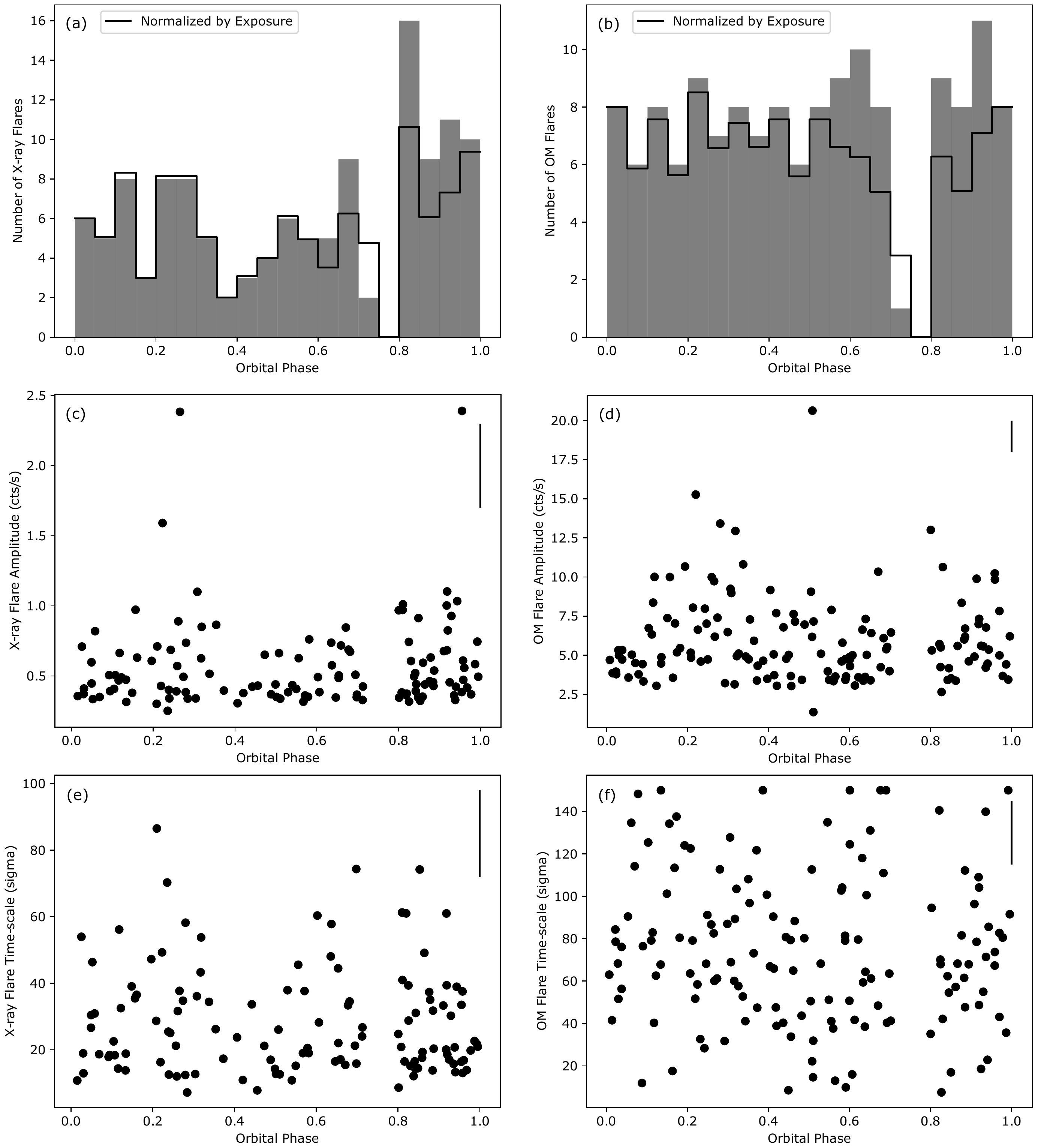}
\caption{From top to bottom, the figures show the total number (a and b), the best-fit Gaussian amplitudes (c and d), and the best-fit Gaussian sigma (e and f) of the flares detected by EPIC (left) and OM (right) at different orbital phases. In the panels (a) and (b), the step function shows the distribution of the number of flares normalized by the exposure of each phase bin w.r.t. that of the first bin. Note that the gaps around phase $0.75$ are due to the X-ray/optical eclipses. The vertical bars at the upper right corners show the typical estimates of the uncertainties. }
\label{fig:phased_flare}
\end{figure*}

\section{X-ray and optical/UV Results}

\subsection{Light curves}
\label{sec:xmm_curve}
\src\ was clearly detected in X-rays (EPIC) and optical/UV (OM), with net count rates of 0.2 cts~s$^{-1}$ (pn + MOS~1/2) and 6 cts~s$^{-1}$, respectively (Figure~\ref{fig:epic_lc} \& \ref{fig:om_lc}). 
Both light curves are dominated by strong flares on time-scales of $\sim10$--$100$~sec. In addition, each light curve shows three eclipses, which perfectly match the orbital ephemeris presented in S16 (Figure \ref{fig:epic_lc} \& \ref{fig:om_lc}). During the eclipse phases, the X-ray and optical fluxes are lower, and the flaring phenomenon is not observed. We tested both light curves for periodic signals longer than 1~sec using \texttt{powspec}, with no clear periodicity found in the frequency range $10^{-5}$~Hz to 1~Hz. 

\begin{figure}
\centering
\includegraphics[width=0.47\textwidth]{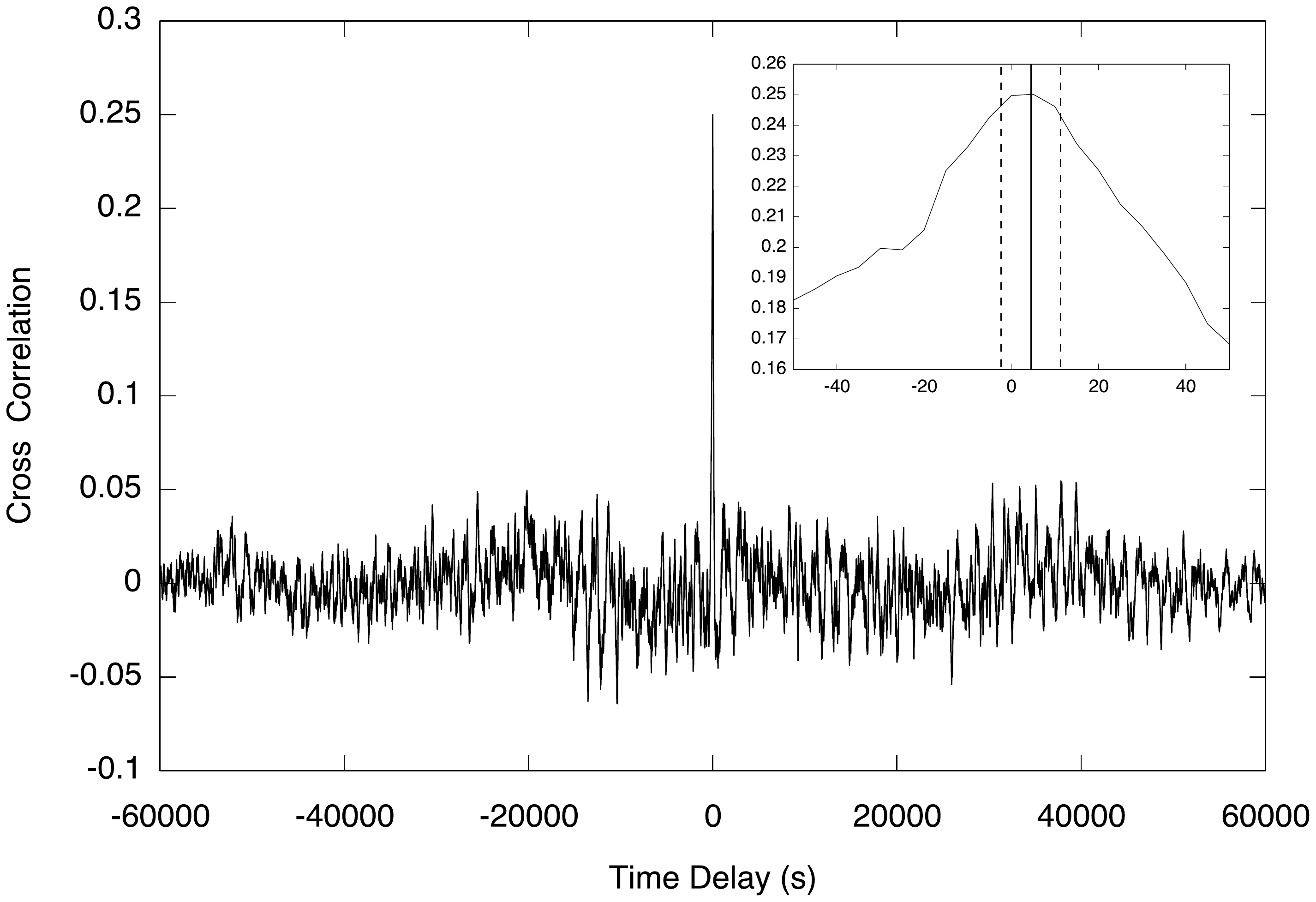}
\caption{The cross-correlation function (5-sec resolution) between the EPIC and the OM light curves computed by \texttt{crosscor}. 
The cross-correlations were normalized by dividing by $\sqrt{N_1N_2}$, where $N_1$ and $N_2$ are the numbers of bins in the two light curves, respectively. 
The three eclipse phases were excluded in the calculation. 
In the function, a positive time delay ($\Delta t_d$) refers an optical/UV emission delay w.r.t. the X-ray emission. 
No evidence of time delay is found in the analysis with $\Delta t_d=4.5\pm6.8$~sec (indicated by the vertical lines in the zoomed-in version in the inset box). 
}
\label{fig:cross}
\end{figure}

\subsubsection{Flare properties}
Here we discuss the detailed properties of the flares.
Except for those X-ray flares detected during the OM observing gaps, almost every X-ray flare has an optical/UV counterpart. The converse is not true: some 
optical/UV flares are not detected in the X-rays (e.g., the flare at $t\approx58000$~sec in Figure \ref{fig:epic_lc}).

As a simple initial model for the flares, we fit every individual flare detected by EPIC and OM with a Gaussian. 
In the fitting, we did not include the data in the eclipses, and simply assumed a constant baseline as the ``quiescent'' emission. 
While the choice of a Gaussian is somewhat arbitrary, it turns out that this model can describe the flares reasonably well (see Figure \ref{fig:epic_lc} \& \ref{fig:om_lc}). 

For the EPIC data, the light curve was binned in 30~sec intervals to obtain a good balance between the signal-to-noise and the timing resolution. Using an initial guess of $0.1$~cts~s$^{-1}$ as the baseline, we iteratively find new flares and re-fit the light curve until no data point exceeds over the model by 2$\sigma$. The same algorithm was also applied for the OM light curve, but a smaller binning factor of 20~sec and a higher initial baseline of $4$~cts~s$^{-1}$ were adopted. 
A higher detection threshold of 4.5$\sigma$ was also used to avoid over-fitting. Otherwise, we found that all residual features were fit with low-amplitude ``flares".
We emphasize that this technique is not designed to achieve a perfect statistical model of the data, but instead to give a first-order sense of the frequency and amplitude of the flares.

As shown in Figures \ref{fig:epic_lc} and \ref{fig:om_lc}, the X-ray and optical/UV flares that are visually obvious are all identified by our technique.
Outside of eclipse, on average we detect an X-ray (optical/UV) flare every 540 (470)~sec.
The occurrence rate is high compared with PSRs~J1023+0038 and J1227$-$4853, of which the flare occurrence rates are just up to a few tens of events per day on average \citep{2013A&A...550A..89D,2015ApJ...806..148B,2018ApJ...858L..12P,2018MNRAS.477.1120K}.
The flaring state of PSR~J1023+0038 observed in optical occupied about 15.6--22\% of the time during the \textit{Kepler K2} observation \citep{2018ApJ...858L..12P,2018MNRAS.477.1120K}. Using the FWHM durations of the detected OM flares, we found that the flaring fraction is $\approx43$\% for \src, which is double the fraction seen in PSR~J1023+0038.
We show a selection of 16 pairs of flares in Figure \ref{fig:16flare}. 

Figure \ref{fig:phased_flare} shows the best-fit parameters for each detected flare as a function of orbital phase. 
The median flare time-scale (defined by the best-fit Gaussian sigma) for the X-ray (optical/UV) flares is 
24~sec (72~sec), equivalent to a FWHM of 57~sec (170~sec).
These time-scales are consistent with some fastest flares seen in the two tMSPs (i.e., duration of a few minutes or less), but still significantly shorter than that of the slowest events observed (i.e., $\approx45$~min -- 14~hours; \citealt{2013A&A...550A..89D,2015ApJ...806..148B, 2018ApJ...858L..12P}). The typical uncertainties for the per flare amplitudes and time-scales for the EPIC (OM) are 0.3~cts~s$^{-1}$ (1.0~cts~s$^{-1}$) and 13~sec (15~sec), though the uncertainties for some flares may be underestimated if the assumed model is a poor fit. This figure also shows that there is no significant evidence for a correlation between flare occurrence or properties with orbital phase. The brightest flares reach $L_X$ (0.2--10 keV) of $\sim 2\times 10^{34}$ erg s$^{-1}$.

To study the relationship between the X-ray and optical/UV flares, we used the \texttt{HEAsoft} task \texttt{crosscor} to calculate the cross-correlation function between the light curves, parameterizing the time delay of the optical/UV emission as $\Delta t_d$. We formally find that the cross-correlation peaks at $\Delta t_d=5$~sec: the X-rays lead the optical/UV emission by about 5 seconds. To determine the uncertainty in this value, we simulated $10^4$ pairs of light curves based on the EPIC and OM data, and repeated the \texttt{crosscor} calculations with the simulated data. The distribution of the simulated cross-correlation peaks (Figure \ref{fig:cross}) gives $\Delta t_d=4.5\pm6.8$~sec (90\% confidence interval). Hence there is no strong evidence for a time delay of the optical/UV emission compared to the X-ray emission. To put this in context, the light curve modeling of S16 found that the outer edge of the accretion disk is $\sim 0.98 R_{\odot}$ ($\sim 2.3$ light-seconds) from the neutron star.

In Figure \ref{fig:sigsig} we directly compare the multi-wavelength properties of 62 optical/UV flares, each of which has one and only one detectable X-ray counterpart within $\pm60$~sec (i.e., equal to two bins of the X-ray light curve shown in Figure \ref{fig:epic_lc}). Interestingly, most of the flares have longer time-scales in optical/UV than in X-rays (e.g., Figure \ref{fig:16flare}h), with only 5 counterexamples found. We checked whether the trend could be affected by the generous $\pm60$~sec matching criterion, by alternatively adopting $\pm10$ and $\pm100$~sec as the allowed time offset, but found similar results.

The average optical-to-X-ray time-scale ratio is 3.5 (with a standard deviation of 2.4), although the samples do not seem to follow a constant ratio (Figure \ref{fig:sigsig}a). For the five outliers, three of them have time-scale ratios larger than 0.8 with the X-ray time-scales just $\lesssim10$~sec longer. Therefore, their time-scale ratios could be consistent, considering the per-fit uncertainties. The other two outlier flares have rather low time-scale ratios ($\lesssim0.7$), and their zoomed-in light curves can be found in Figure \ref{fig:16flare}o and \ref{fig:16flare}e. The X-ray and optical/UV flare amplitudes are roughly correlated (larger X-ray flares generally lead to larger optical/UV flares), but with a large scatter (Figure \ref{fig:sigsig}b). The amplitudes themselves arguably change on short time-scales (e.g., Figure \ref{fig:16flare}f), which could explain part of this scatter.

\begin{figure}
\centering
\includegraphics[width=0.47\textwidth]{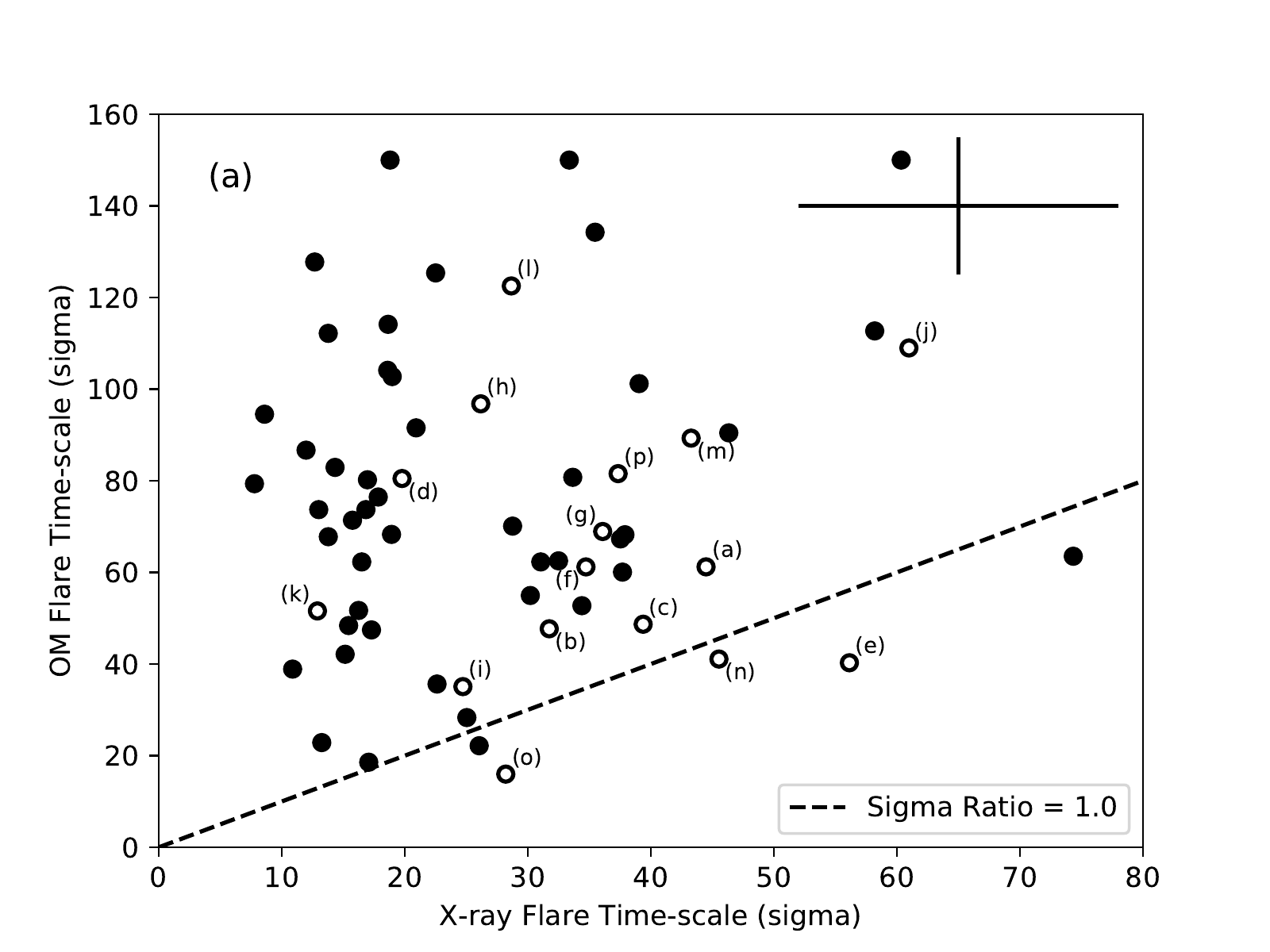}
\includegraphics[width=0.47\textwidth]{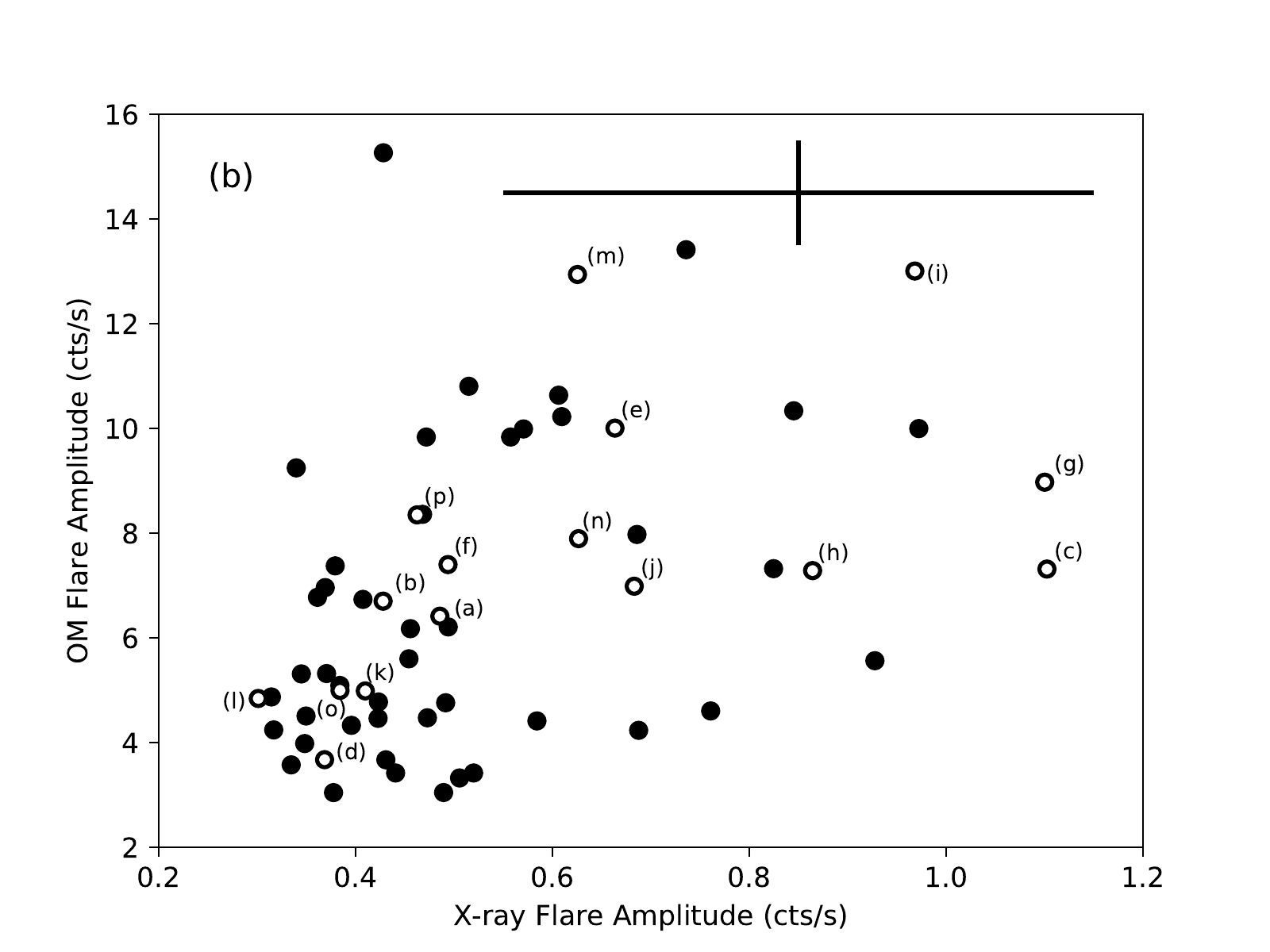}
\caption{The figures show the comparisons between the flare characteristics observed in X-ray and optical/UV. 
The upper panel (a) refers to the flare time-scale comparison, while the lower panel (b) is for the flare amplitudes. 
In general, the flares last significantly longer in optical/UV than in X-rays (see the dashed line, which indicates the time-scale ratio as one). No clear relation can be found between the optical/UV and X-ray amplitudes. The 16 representative flares that are shown in Figure \ref{fig:16flare} are indicated by empty circles and labelled accordingly from (a) through (p). The crosses at the upper right corners show the typical estimates of the uncertainties. }
\label{fig:sigsig}
\end{figure}

\subsection{X-ray Spectral analysis}
\label{sec:spec_analy}
Since the X-ray data are strongly variable, we separated the data into several groups for independent spectral analyses: 

\begin{itemize}\setlength\itemsep{0.0cm}
\item \textit{Average1} (MOS~1/2): the out-of-eclipse spectrum, selected based on the orbital timing solution in S16. 
\item \textit{High1} (MOS~1/2): a subset of \textit{Average1}, with EPIC net count rates $>0.6$ cts~s$^{-1}$ per 30-sec bin (Figure \ref{fig:epic_lc}). 
\item \textit{Medium1} (MOS~1/2): similar with \textit{High1}, but with count rates between $0.3$ and 0.6 cts~s$^{-1}$. 
\item \textit{Low1} (MOS~1/2): similar with \textit{High1}, but with count rates $\leqslant0.3$ cts~s$^{-1}$. 
\item \textit{Eclipse} (pn and MOS~1/2): data taken during the three eclipse phases. 
\end{itemize}

As mentioned in \S\ref{sec:xmm_re}, the pn and MOS~1/2 data have different time coverages after the flaring background filtering, in which the MOS~1/2 data are less affected. Therefore, only the MOS~1/2 spectra were used in the analysis, except for the \textit{Eclipse} group, in which the pn data was included. 
However, we also prepared four additional data groups with suffix ``\textit{2}'' (e.g., \textit{Average2}), which contain also the pn spectra. However, the fitting results obtained from these datasets are for reference purposes only, and the following discussions are all based on the primary datasets (suffix ``\textit{1}''). 

All the X-ray spectral fitting processes were done by \texttt{XSPEC} (version 12.9.1m). 
The spectra were binned using \texttt{grppha} to at least 20 counts per bin, and then fit with given spectral models based on chi-squared statistic ($\chi^2$), unless otherwise stated. 
Various spectral models have been tested. 
In these models, two components of hydrogen column density (\nh) were assumed for the Galactic foreground and intrinsic absorptions, respectively\footnote{Except for the \textit{Eclipse} data, as the resultant spectrum does not have a sufficient S/N for modelling an intrinsic X-ray absorption. }. The Galactic value was fixed to \nh\ $=3.66\times10^{20}$\cm\ obtained from the Leiden/Argentine/Bonn map \citep{2005A&A...440..775K}, while the intrinsic \nh\ was allowed to vary. Besides, an energy-independent multiplicative factor was employed to the whole physical model to account for the cross-calibrations among the EPIC detectors (see the footnotes in Table \ref{tab:xmm_spec} for the definitions). 
Throughout the analysis, all the listed uncertainties were calculated at 90\% confidence level for one interesting parameter (i.e., $\Delta\chi^2=2.71$; \citealt{1976ApJ...210..642A}). 

\begin{figure*}
\centering
\includegraphics[width=0.8\textwidth]{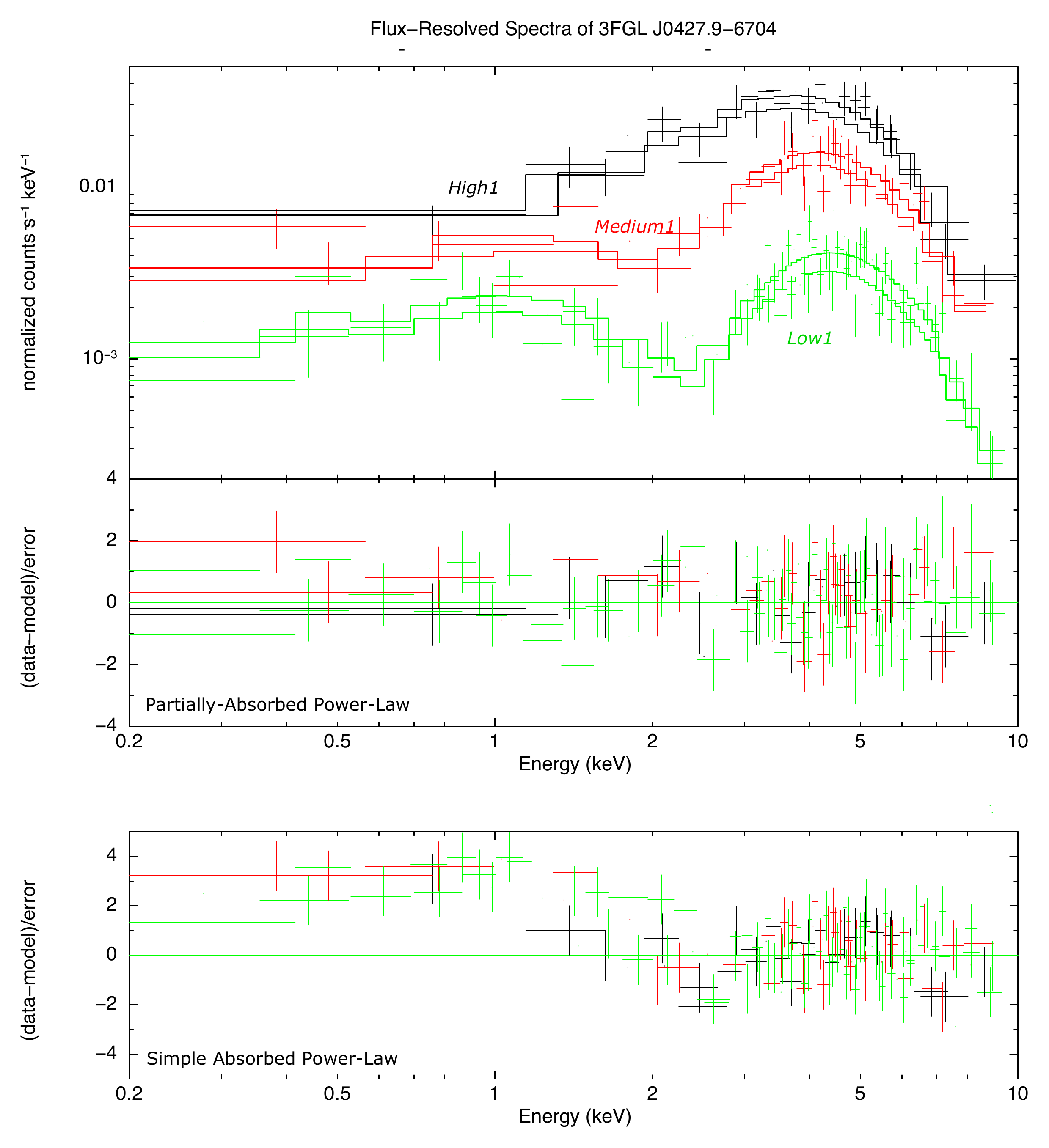}
\caption{Upper: The \textit{XMM-Newton} (MOS~1/2) phase-resolved spectra of \src\ with the best-fit partially-absorbed power-law models (see Table \ref{tab:xmm_spec} for the best-fit parameters). From bright to faint, the spectra are classified into \textit{High1} ($>0.6$~cts~s$^{-1}$; black), \textit{Medium1} (0.3--0.6~cts~s$^{-1}$; red), and \textit{Low1} ($<0.3$~cts~s$^{-1}$; green). The detailed scheme of classification can be found in \S\ref{sec:spec_analy}. Lower: A residual plot with respect to the best-fit simple absorbed power-law models.}
\label{fig:epic_spec}
\end{figure*}

\subsubsection{Average spectra}
We first tried a simple power-law to gain insights into detailed spectral modelling. 
The best-fit gives a bad fitting statistic of $\chi_\nu^2=2.4$ (188 d.o.f.) with a very hard photon index of $\Gamma\approx0.5$ and an intrinsic absorption of \nh\ $\approx4\times10^{22}$\cm. 
In general, the spectral shape is saddle-like, with a soft X-ray excess below 2~keV, which is the primary cause of the bad fit. 
An additional thermal emission component corresponding to the soft excess was thus considered. 
Given that the huge intrinsic absorption for the power-law component would have absorbed most of the soft thermal X-rays (if any), we once assumed an independent intrinsic absorption \textit{only} for the thermal emission (it would be the case if, e.g., the emission regions for the thermal and non-thermal components are different). However, the best-fit value of this extra \nh\ always goes to zero (e.g., \nh\ $<2\times10^{20}$\cm\ for a blackbody fit with the best-fit value found at zero; the same situation was also seen in all other data groups). We therefore removed this absorption component and left the thermal emission solely absorbed by the Galactic foreground medium. 

Statistically either a single temperature blackbody (\texttt{bbodyrad} in \texttt{XSPEC}) or a multi-temperature disk blackbody (\texttt{diskbb}) can improve the power-law fit significantly.
The \texttt{diskbb} model is slightly better in terms of $\chi^2$ statistics ($\chi^2_\nu=1.17$ and 1.09 for \texttt{bbodyrad} and \texttt{diskbb}, respectively). 
With a soft thermal component, the best-fit photon index becomes softer (i.e., $\Gamma\approx0.9$--1.1), and comparable to other redback MSP systems (see, e.g., \citealt{2018ApJ...864...23L}). 
As expected, the thermal component is very soft with best-fit temperatures of $T_{\rm bb}\approx0.3$~keV for \texttt{bbodyrad} and $T_{\rm in}\approx0.5$~keV (the inner disk temperature) for \texttt{diskbb}. 
Comparing with the non-thermal X-ray component, they are very faint. We use the best-fit normalizations and the distance of $d=2.3$~kpc estimated by \textit{Gaia} \citep{2018AJ....156...58B}, which is consistent with the distance from modeling the light curve in S16. The best-fit blackbody emission region is $R_{\rm bb}\approx0.2$~km in radius, and the inner disk radius of the disk is $R_{\rm in}\approx0.1$~km ($\cos i=0.18$ assumed; S16). Apparently, the inner radius is much smaller than the size of a typical neutron star (radius of $\sim10$~km), making the \texttt{diskbb} case highly unlikely. We also fit the spectrum with the neutron star atmospheric model, \texttt{nsa}, for more realistic estimates on the temperature and the emission size. 
In the \texttt{nsa} model, we assumed a non-magnetized (i.e., $B<10^9$~G) neutron star of $M_{\rm NS}=1.4M_\sun$ and $R_{\rm NS}=10$~km.
The best-fit temperature is lower ($T_{\rm nsa}\approx0.2$~keV) and the emission region is significantly bigger ($R_{\rm nsa}\approx0.5$~km). 

Besides an additional thermal component, a partial covering fraction absorption (\texttt{pcfabs}) with a simple power-law (hereafter called a partially-absorbed power-law) can also explain the saddle-shaped X-ray spectrum very well ($\chi^2_\nu=1.1$ with 187 d.o.f.; Figure \ref{fig:epic_spec}). 
In the best-fit partially-absorbed power-law, the intrinsic absorber only covers about 97\% of the X-rays in our line-of-sight, and 3\% of the X-ray emission ``leaked'' to produce the observed soft X-ray bump. 
Comparing with the simple power-law models with/without a thermal component, the intrinsic hydrogen column density is significantly higher (\nh\ $=(10.9\pm0.5)\times10^{22}$\cm). The photon index is also much softer (still hard though; $\Gamma=1.44\pm0.05$) and closer to that of PSRs~J1023+0038 and J1227$-$4853 during the sub-luminous disk state, which are typically around $\Gamma\approx1.6$--1.8 (e.g., \citealt{2013A&A...550A..89D,2014ApJ...797..111L,2015ApJ...806..148B}).
These high intrinsic \nh\ and $\Gamma$ values lead the inferred non-thermal X-ray luminosity (0.2--10~keV) to $L_{\rm nth}=(1.75\pm0.05)\times 10^{33}$\lum, which is also the highest among all the models. 
Despite the highest degree-of-freedom among the fits, the partially-absorbed power-law fit has the best performance in terms of $\chi^2$ statistic, indicating that it is a better model for the \textit{Average1} spectrum statistically. 

In the 84-ksec \textit{NuSTAR} observation taken in 2016 May (about 1 year before the \textit{XMM-Newton} observation), the photon index ($\Gamma_{\rm nu}=1.68^{+0.09}_{-0.08}$; S16) is much softer than any of the photon indices obtained in \textit{Average1}. 
The photon index deviation could be due to either (i) yearly spectral variability of \src, or (ii) a high-energy exponential cut-off at $\gtrsim10$~keV. 
For testing the latter case, we performed a joint spectral fitting of the \textit{Average1} and \textit{NuSTAR} (obtained from S16) spectra with \texttt{constant*phabs*pcfabs*cutoffpl}, where \texttt{cutoffpl} can be written as $\frac{dN}{dE}\propto E^{-\Gamma}\exp (-\frac{E}{E_{\rm c}})$. 
We fixed the \texttt{pcfabs} absorption component at the best-fit parameters obtained from \textit{Average1} for simplicity, and found that a power-law model with $E_{\rm c}\approx30$~keV can fit the joint \textit{XMM-Newton}-\textit{NuSTAR} spectrum very well, yielding a fitting statistic of $\chi_\nu^2=1.1$ (471 d.o.f.). 
However, we noticed that \src\ was about 40\% brighter in the \textit{NuSTAR} observation (see the ``\textit{X}+\textit{N}'' row in Table \ref{tab:xmm_spec}), strongly suggesting \src\ as a long-term X-ray variable (e.g., the flare occurrence rate changes over a yearly time-scale). Therefore, the observed high-energy cut-off may not be significant. 

\subsubsection{Flux-resolved spectra}
Like \textit{Average1}, the three flux-resolved spectra are all saddle-like. Nevertheless, their spectral shapes are slightly different from each other. From \textit{Low1} to \textit{High1}, the saddle-shaped feature becomes less obvious (Figure \ref{fig:epic_spec}). 
With the same spectral models applied on \textit{Average1}, we characterized the spectral features of these groups. Except the simple absorbed power-law model, all the applied spectral models are equally good for the data statistically\footnote{In \textit{High1}, the \texttt{diskbb}, \texttt{bbody}, and \texttt{nsa} temperatures were all fixed to the best-fit values obtained from $Average$. Otherwise, the fits do not converge.}. 

Similar to \textit{Average1}, the inner disk radii inferred are all too small for a neutron star system. 
For the model composed of \texttt{power} and \texttt{bbodyrad}, the non-thermal X-ray emission still dominates the entire energy band. Although the photon index does not significantly change, the intrinsic \nh\ for the power-law component drops dramatically as the non-thermal X-ray luminosity increases (i.e., \nh/$10^{22}$\cm: $9.2\Rightarrow6.3\Rightarrow4.1$ as $L_{\rm nth}$/$10^{33}$\lum: $0.8\Rightarrow2.9\Rightarrow5.4$). For the blackbody component, the temperature stays around 0.2--0.3~keV. The inferred emission size varies from group to group, but the changes are not significant as the uncertainties are high, e.g., all the sizes can be consistent with $R_{\rm bb}\approx0.2$~km. 
Very similar results can be found in the \texttt{nsa} fits, with lower temperatures ($T_{\rm nsa}\approx0.1$--$0.2$~keV) and larger emission regions ($R_{\rm nsa}\approx0.5$--$1.5$~km). 

For the partially-absorbed power-law fits, the \nh\ decreasing trend aforementioned is still observed, however, with higher variability (i.e., \nh/$10^{22}$\cm: $15\Rightarrow10\Rightarrow4.9$ as $L_{\rm nth}$/$10^{33}$\lum: $1.6\Rightarrow4.5\Rightarrow5.7$). The photon index also becomes harder as the X-ray luminosity goes up (i.e., $\Gamma$: $1.8\Rightarrow1.6\Rightarrow1.2$). 
Surprisingly, the coverage of the X-ray absorber does not change significantly among the flux-resolved groups ($\gtrsim$ 90\% roughly). 
To further check whether this relation holds at lower luminosities, we extracted an extra set of spectra with MOS~1/2 data of $<$0.1~cts~s$^{-1}$ (labelled as \textit{CR$<$0.1} in Table \ref{tab:xmm_spec}), which confirms the speculation with an even higher column density of \nh\ $\approx2\times10^{23}$\cm. 

\begin{table*}
\renewcommand{\arraystretch}{1.5}
\scriptsize
\centering 
\caption{X-ray spectral properties of \src}
\begin{tabular}{l@{\hskip 0.1cm}l@{\hskip -0.4cm}rcccccrccr}
\toprule
Dataset & Model$^{\rm a}$ & \multicolumn{1}{c}{\nh} & $C_1$$^{\rm b}$ & $C_2$$^{\rm b}$ & Fraction & $\Gamma$ & \multicolumn{1}{c}{$T_{}$ or $E_{\rm c}$} & \multicolumn{1}{c}{Radius} & $L_{\rm nth}$ & $\frac{L_{\rm nth}}{L_{\rm tot}}$ & \multicolumn{1}{c}{$\chi_\nu^2$} \\
 & & \multicolumn{1}{c}{($10^{22}$\cm)} & & & (\%) & & \multicolumn{1}{c}{(keV)} & \multicolumn{1}{c}{(meters)} & ($10^{33}$\lum) & (\%) & \\
\hline

\textit{Average1} & \texttt{pcfabs*pow} & $10.9^{+0.5}_{-0.5}$ & \nodata\ & $1.1^{+0.1}_{-0.1}$ & $96.5^{+1.0}_{-1.2}$ & $1.44^{+0.05}_{-0.05}$ & \nodata\ & \nodata\ & $1.75^{+0.05}_{-0.05}$ & \nodata\ & 197.7/187 \\

 & \texttt{phabs*pow} & $3.8^{+1.0}_{-1.0}$ & \nodata\ & $1.0^{+0.1}_{-0.1}$ & \nodata\ & $0.48^{+0.22}_{-0.22}$ & \nodata\ & \nodata\ & $1.02^{+0.08}_{-0.07}$ & \nodata\ & 450.3/188 \\

 & \texttt{diskbb+phabs*pow} & $7.7^{+2.0}_{-1.6}$ & \nodata\ & $1.1^{+0.1}_{-0.1}$ & \nodata\ & $1.07^{+0.28}_{-0.26}$ & $0.52^{+0.23}_{-0.12}$ & $84^{+55}_{-41}$ & $1.32^{+0.29}_{-0.17}$ & 98 & 203.3/186 \\

 & \texttt{bbody+phabs*pow} & $6.5^{+1.5}_{-1.2}$ & \nodata\ & $1.0^{+0.1}_{-0.1}$ & \nodata\ & $0.91^{+0.26}_{-0.23}$ & $0.27^{+0.05}_{-0.04}$ & $154^{+56}_{-46}$ & $1.20^{+0.19}_{-0.12}$ & 99 & 217.3/186 \\
 

 & \texttt{nsa+phabs*pow}$^{\rm c}$ & $6.8^{+1.4}_{-1.3}$ & \nodata\ & $1.0^{+0.1}_{-0.1}$ & \nodata\ & $0.95^{+0.25}_{-0.24}$ & $0.18^{+0.05}_{-0.04}$ & $507^{+328}_{-195}$ & $1.22^{+0.19}_{-0.13}$ & 99 & 211.0/186 \\

\hline

\textit{Low1} & \texttt{pcfabs*pow} & $15.4^{+0.9}_{-0.8}$ & \nodata\ & $1.1^{+0.1}_{-0.1}$ & $97.9^{+0.8}_{-1.2}$ & $1.75^{+0.05}_{-0.05}$ & \nodata\ & \nodata\ & $1.61^{+0.07}_{-0.07}$ & \nodata\ & 114.4/95 \\

 & \texttt{phabs*pow} & $6.9^{+2.7}_{-2.3}$ & \nodata\ & $1.1^{+0.1}_{-0.1}$ & \nodata\ & $0.75^{+0.43}_{-0.39}$ & \nodata\ & \nodata\ & $0.71^{+0.18}_{-0.10}$ & \nodata\ & 267.4/96 \\

 & \texttt{diskbb+phabs*pow} & $10.4^{+2.9}_{-2.5}$ & \nodata\ & $1.1^{+0.1}_{-0.1}$ & \nodata\ & $1.17^{+0.43}_{-0.39}$ & $0.42^{+0.13}_{-0.08}$ & $113^{+64}_{-47}$ & $0.90^{+0.40}_{-0.18}$ & 98 & 113.2/94 \\

 & \texttt{bbody+phabs*pow} & $9.2^{+2.7}_{-2.3}$ & \nodata\ & $1.1^{+0.1}_{-0.1}$ & \nodata\ & $1.04^{+0.42}_{-0.38}$ & $0.24^{+0.04}_{-0.03}$ & $170^{+61}_{-49}$ & $0.82^{+0.29}_{-0.14}$ & 98 & 118.2/94 \\
 
 & \texttt{nsa+phabs*pow} & $9.8^{+2.7}_{-2.4}$ & \nodata\ & $1.1^{+0.1}_{-0.1}$ & \nodata\ & $1.11^{+0.42}_{-0.38}$ & $0.16^{+0.04}_{-0.03}$ & $581^{+366}_{-226}$ & $0.86^{+0.33}_{-0.16}$ & 98 & 115.0/94 \\

\hline

\textit{Medium1} & \texttt{pcfabs*pow} & $10.4^{+0.8}_{-0.7}$ & \nodata\ & $1.0^{+0.1}_{-0.1}$ & $98.0^{+0.9}_{-1.8}$ & $1.60^{+0.07}_{-0.08}$ & \nodata\ & \nodata\ & $4.53^{+0.25}_{-0.25}$ & \nodata\ & 51.0/48 \\
 
 & \texttt{phabs*pow} & $5.1^{+1.8}_{-1.6}$ & \nodata\ & $1.0^{+0.1}_{-0.1}$ & \nodata\ & $0.77^{+0.41}_{-0.39}$ & \nodata\ & \nodata\ & $2.59^{+0.55}_{-0.34}$ & \nodata\ & 112.7/49 \\

 & \texttt{diskbb+phabs*pow} & $6.8^{+2.4}_{-1.8}$ & \nodata\ & $1.0^{+0.1}_{-0.1}$ & \nodata\ & $1.05^{+0.45}_{-0.39}$ & $0.34^{+0.23}_{-0.09}$ & $269^{+301}_{-178}$ & $2.98^{+1.08}_{-0.49}$ & 99 & 49.5/47 \\

 & \texttt{bbody+phabs*pow} & $6.3^{+2.1}_{-1.7}$ & \nodata\ & $1.0^{+0.1}_{-0.1}$ & \nodata\ & $0.98^{+0.42}_{-0.38}$ & $0.21^{+0.07}_{-0.04}$ & $374^{+266}_{-182}$ & $2.86^{+0.84}_{-0.43}$ & 99 & 51.9/47 \\
 
 & \texttt{nsa+phabs*pow} & $6.6^{+2.1}_{-1.7}$ & \nodata\ & $1.0^{+0.1}_{-0.1}$ & \nodata\ & $1.02^{+0.42}_{-0.39}$ & $0.12^{+0.06}_{-0.04}$ & $1550^{+2162}_{-943}$ & $2.93^{+0.92}_{-0.46}$ & 99 & 50.1/47 \\

\hline

\textit{High1} & \texttt{pcfabs*pow} & $4.9^{+0.6}_{-0.5}$ & \nodata\ & $1.0^{+0.1}_{-0.1}$ & $95.7^{+2.2}_{-3.4}$ & $1.22^{+0.12}_{-0.13}$ & \nodata\ & \nodata\ & $5.69^{+0.36}_{-0.36}$ & \nodata\ & 21.9/33 \\

 & \texttt{phabs*pow} & $2.7^{+1.1}_{-0.8}$ & \nodata\ & $1.0^{+0.1}_{-0.1}$ & \nodata\ & $0.86^{+0.34}_{-0.31}$ & \nodata\ & \nodata\ & $4.63^{+0.80}_{-0.59}$ & \nodata\ & 46.9/34 \\

 & \texttt{diskbb+phabs*pow} & $4.4^{+1.6}_{-1.2}$ & \nodata\ & $1.0^{+0.1}_{-0.1}$ & \nodata\ & $1.20^{+0.40}_{-0.36}$ & (0.52)$^{\rm d}$ & $144^{+22}_{-26}$ & $5.55^{+2.00}_{-0.97}$ & 99 & 22.9/33 \\

 & \texttt{bbody+phabs*pow} & $4.1^{+1.5}_{-1.1}$ & \nodata\ & $1.0^{+0.1}_{-0.1}$ & \nodata\ & $1.16^{+0.38}_{-0.34}$ & (0.27)$^{\rm d}$ & $264^{+41}_{-49}$ & $5.38^{+1.66}_{-0.87}$ & 99 & 23.2/33 \\
 
 & \texttt{nsa+phabs*pow} & $4.1^{+1.5}_{-1.1}$ & \nodata\ & $1.0^{+0.1}_{-0.1}$ & \nodata\ & $1.15^{+0.39}_{-0.34}$ & (0.18)$^{\rm d}$ & $855^{+134}_{-158}$ & $5.37^{+1.67}_{-0.87}$ & 99 & 23.4/33 \\

\hline
\hline


\textit{Average2} & \texttt{pcfabs*pow} & $10.4^{+0.4}_{-0.4}$ & $1.0^{+0.1}_{-0.1}$ & $1.0^{+0.1}_{-0.1}$ & $96.5^{+0.8}_{-0.9}$ & $1.43^{+0.04}_{-0.04}$ & \nodata\ & \nodata\ & $1.74^{+0.04}_{-0.04}$ & \nodata\ & 320.0/300 \\

 & \texttt{phabs*pow} & $4.3^{+0.8}_{-0.7}$ & $1.0^{+0.1}_{-0.1}$ & $1.0^{+0.1}_{-0.1}$ & \nodata\ & $0.69^{+0.16}_{-0.16}$ & \nodata\ & \nodata\ & $1.03^{+0.08}_{-0.07}$ & \nodata\ & 750.0/301 \\

 & \texttt{diskbb+phabs*pow} & $7.9^{+1.4}_{-1.2}$ & $1.0^{+0.1}_{-0.1}$ & $1.0^{+0.1}_{-0.1}$ & \nodata\ & $1.18^{+0.20}_{-0.18}$ & $0.54^{+0.15}_{-0.09}$ & $78^{+34}_{-28}$ & $1.39^{+0.25}_{-0.17}$ & 98 & 327.9/299 \\

 & \texttt{bbody+phabs*pow} & $6.7^{+1.1}_{-0.9}$ & $1.0^{+0.1}_{-0.1}$ & $1.0^{+0.1}_{-0.1}$ & \nodata\ & $1.03^{+0.18}_{-0.17}$ & $0.27^{+0.03}_{-0.03}$ & $156^{+36}_{-31}$ & $1.25^{+0.16}_{-0.12}$ & 99 & 347.6/299 \\
 
 & \texttt{nsa+phabs*pow} & $7.0^{+1.0}_{-1.0}$ & $1.0^{+0.1}_{-0.1}$ & $1.0^{+0.1}_{-0.1}$ & \nodata\ & $1.07^{+0.18}_{-0.17}$ & $0.18^{+0.03}_{-0.03}$ & $476^{+186}_{-126}$ & $1.28^{+0.17}_{-0.13}$ & 99 & 338.5/299 \\
 
\hline

\textit{Low2} & \texttt{pcfabs*pow} & $15.4^{+0.7}_{-0.7}$ & $1.0^{+0.1}_{-0.1}$ & $1.1^{+0.1}_{-0.1}$ & $97.7^{+0.6}_{-0.9}$ & $1.74^{+0.04}_{-0.04}$ & \nodata\ & \nodata\ & $1.58^{+0.06}_{-0.06}$ & \nodata\ & 171.4/154 \\
 
 & \texttt{phabs*pow} & $7.2^{+2.0}_{-1.8}$ & $1.1^{+0.1}_{-0.1}$ & $1.1^{+0.1}_{-0.1}$ & \nodata\ & $0.85^{+0.31}_{-0.29}$ & \nodata\ & \nodata\ & $0.69^{+0.15}_{-0.10}$ & \nodata\ & 455.5/155 \\

 & \texttt{diskbb+phabs*pow} & $10.6^{+2.2}_{-2.0}$ & $1.0^{+0.1}_{-0.1}$ & $1.1^{+0.1}_{-0.1}$ & \nodata\ & $1.24^{+0.32}_{-0.29}$ & $0.42^{+0.08}_{-0.06}$ & $112^{+42}_{-34}$ & $0.92^{+0.31}_{-0.17}$ & 98 & 168.3/153 \\

 & \texttt{bbody+phabs*pow} & $9.4^{+2.1}_{-1.8}$ & $1.0^{+0.1}_{-0.1}$ & $1.1^{+0.1}_{-0.1}$ & \nodata\ & $1.12^{+0.31}_{-0.28}$ & $0.24^{+0.03}_{-0.02}$ & $175^{+42}_{-36}$ & $0.83^{+0.23}_{-0.14}$ & 98 & 175.9/153 \\
 
 & \texttt{nsa+phabs*pow} & $10.0^{+2.1}_{-1.9}$ & $1.0^{+0.1}_{-0.1}$ & $1.1^{+0.1}_{-0.1}$ & \nodata\ & $1.18^{+0.30}_{-0.29}$ & $0.16^{+0.03}_{-0.02}$ & $579^{+238}_{-161}$ & $0.87^{+0.26}_{-0.15}$ & 98 & 170.7/153 \\

\hline

\textit{Medium2} & \texttt{pcfabs*pow} & $10.3^{+0.6}_{-0.6}$ & $1.0^{+0.1}_{-0.1}$ & $1.0^{+0.1}_{-0.1}$ & $98.2^{+0.7}_{-1.0}$ & $1.58^{+0.06}_{-0.06}$ & \nodata\ & \nodata\ & $4.62^{+0.20}_{-0.20}$ & \nodata\ & 73.5/79 \\
 
 & \texttt{phabs*pow} & $6.2^{+1.5}_{-1.3}$ & $1.0^{+0.1}_{-0.1}$ & $1.0^{+0.1}_{-0.1}$ & \nodata\ & $1.07^{+0.30}_{-0.28}$ & \nodata\ & \nodata\ & $2.92^{+0.71}_{-0.44}$ & \nodata\ & 181.3/80 \\

 & \texttt{diskbb+phabs*pow} & $8.2^{+2.1}_{-1.6}$ & $1.0^{+0.1}_{-0.1}$ & $1.0^{+0.1}_{-0.1}$ & \nodata\ & $1.32^{+0.33}_{-0.30}$ & $0.43^{+0.28}_{-0.12}$ & $153^{+144}_{-94}$ & $3.54^{+1.41}_{-0.71}$ & 99 & 75.9/78 \\

 & \texttt{bbody+phabs*pow} & $7.4^{+1.7}_{-1.4}$ & $1.0^{+0.1}_{-0.1}$ & $1.0^{+0.1}_{-0.1}$ & \nodata\ & $1.23^{+0.31}_{-0.29}$ & $0.23^{+0.07}_{-0.04}$ & $267^{+140}_{-107}$ & $3.27^{+1.03}_{-0.58}$ & 99 & 80.8/78 \\
 
 & \texttt{nsa+phabs*pow} & $7.7^{+1.7}_{-1.5}$ & $1.0^{+0.1}_{-0.1}$ & $1.0^{+0.1}_{-0.1}$ & \nodata\ & $1.26^{+0.31}_{-0.29}$ & $0.15^{+0.06}_{-0.04}$ & $939^{+915}_{-469}$ & $3.36^{+1.10}_{-0.62}$ & 99 & 78.0/78 \\

\hline

\textit{High2} & \texttt{pcfabs*pow} & $3.7^{+0.4}_{-0.3}$ & $1.1^{+0.1}_{-0.1}$ & $1.1^{+0.1}_{-0.1}$ & $95.4^{+1.9}_{-3.0}$ & $1.02^{+0.10}_{-0.10}$ & \nodata\ & \nodata\ & $4.58^{+0.23}_{-0.23}$ & \nodata\ & 40.5/56 \\

 & \texttt{phabs*pow} & $2.3^{+0.7}_{-0.6}$ & $1.1^{+0.1}_{-0.1}$ & $1.1^{+0.1}_{-0.1}$ & \nodata\ & $0.78^{+0.25}_{-0.24}$ & \nodata\ & \nodata\ & $4.14^{+0.47}_{-0.40}$ & \nodata\ & 69.1/57 \\

 & \texttt{diskbb+phabs*pow} & $3.4^{+0.9}_{-0.8}$ & $1.1^{+0.1}_{-0.1}$ & $1.1^{+0.1}_{-0.1}$ & \nodata\ & $1.00^{+0.27}_{-0.25}$ & (0.54)$^{\rm d}$ & $106^{+15}_{-18}$ & $4.50^{+0.72}_{-0.51}$ & 99 & 41.9/56 \\

 & \texttt{bbody+phabs*pow} & $5.7^{+7.9}_{-1.9}$ & $1.1^{+0.1}_{-0.1}$ & $1.1^{+0.1}_{-0.1}$ & \nodata\ & $1.10^{+2.68}_{-0.40}$ & $1.25^{+1.85}_{-0.98}$ & $40^{+143}_{-19}$ & $4.36^{+2.46}_{-1.88}$ & 90 & 37.9/55 \\
 
 & \texttt{nsa+phabs*pow} & $3.2^{+0.9}_{-0.7}$ & $1.1^{+0.1}_{-0.1}$ & $1.1^{+0.1}_{-0.1}$ & \nodata\ & $0.98^{+0.27}_{-0.25}$ & (0.18)$^{\rm d}$ & $638^{+92}_{-108}$ & $4.46^{+0.68}_{-0.49}$ & 99 & 42.0/56 \\

\hline
\hline


\textit{Eclipse}${\rm ^e}$ & \texttt{pow} & \nodata\ & (1.0) & (1.0) & \nodata\ & $1.90^{+0.45}_{-0.43}$ & \nodata\ & \nodata\ & $0.03^{+0.01}_{-0.01}$ & \nodata\ & \nodata\ \\

\hline

\textit{X}+\textit{N}${\rm ^f}$ & \texttt{pcfabs*cutoffpl} & $(10.9)$ & $1.4^{+0.1}_{-0.1}$ & $1.4^{+0.1}_{-0.1}$ & $(96.5)$ & $1.43^{+0.07}_{-0.07}$ & $31^{+13}_{-7}$ & \nodata\ & $1.76^{+0.07}_{-0.07}$ & \nodata\ & 502.8/471 \\

\hline

\textit{CR$<$0.1} & \texttt{pcfabs*pow} & $20.8^{+3.0}_{-2.4}$ & \nodata & $1.1^{+0.2}_{-0.2}$ & $98.2^{+1.4}_{-4.2}$ & $1.75^{+0.13}_{-0.13}$ & \nodata\ & \nodata\ & $0.91^{+0.11}_{-0.11}$ & \nodata\ & 17.0/18 \\
\hline


\end{tabular}
\footnotetext{ The models are listed in \texttt{XSPEC} parlance. The cross-calibration factor(s) for different instruments and the Galactic absorption are not shown for simplicity, and the complete model forms should therefore read \texttt{constant*phabs*(<Model>)}, where \nh\ $=3.66\times10^{20}$\cm\ (fixed) for the Galactic \texttt{phabs} component. The luminosities were computed in the energy range of 0.2--10~keV with $d=2.3$~kpc. }
\footnotetext{ Except for \textit{X}+\textit{N}, $C_1$ (or $C_2$) is the cross-calibration factor for MOS~1 (MOS~2) w.r.t. pn. (MOS~1 or pn)}
\footnotetext{ For \texttt{nsa}, a non-magnetized (i.e., $B<10^9$~G) neutron star of $M_{\rm NS}=1.4M_\sun$ and $R_{\rm NS}=10$~km is assumed. }
\footnotetext{ The temperatures that could not be well converged were fixed to the values obtained from $Average/Average2$. }
\footnotetext{ As the data quality is low, the cross-calibration factors were both fixed to 1. \textit{C} statistic was also applied ($C=124.0$ with 132 d.o.f.). }
\footnotetext{ \textit{X}+\textit{N} is the \textit{Average1} and \textit{NuSTAR} (data taken from S16) joint fit, where $C_1$ and $C_2$ are the cross-calibration factors for the focal plane modules FPMA an FPMB w.r.t. MOS~1, respectively. The intrinsic \nh, the cross-calibration factor for MOS~2, and the Fraction parameter for \texttt{pcfabs} were all fixed to the values obtained from \textit{Average1} for simplicity. }
\label{tab:xmm_spec}
\end{table*}

\subsubsection{Eclipse spectra}
With the orbital solution presented in S16, we selected data collected in the three eclipse intervals for the following analysis. In addition, we excluded data within the first and last 200 seconds of the eclipses to avoid any residual emission in the eclipse ingress and egress (e.g., a possible mini-flare is seen in the last $\sim100$ seconds of the third eclipse). 

\begin{figure*}
\centering
\includegraphics[width=\textwidth]{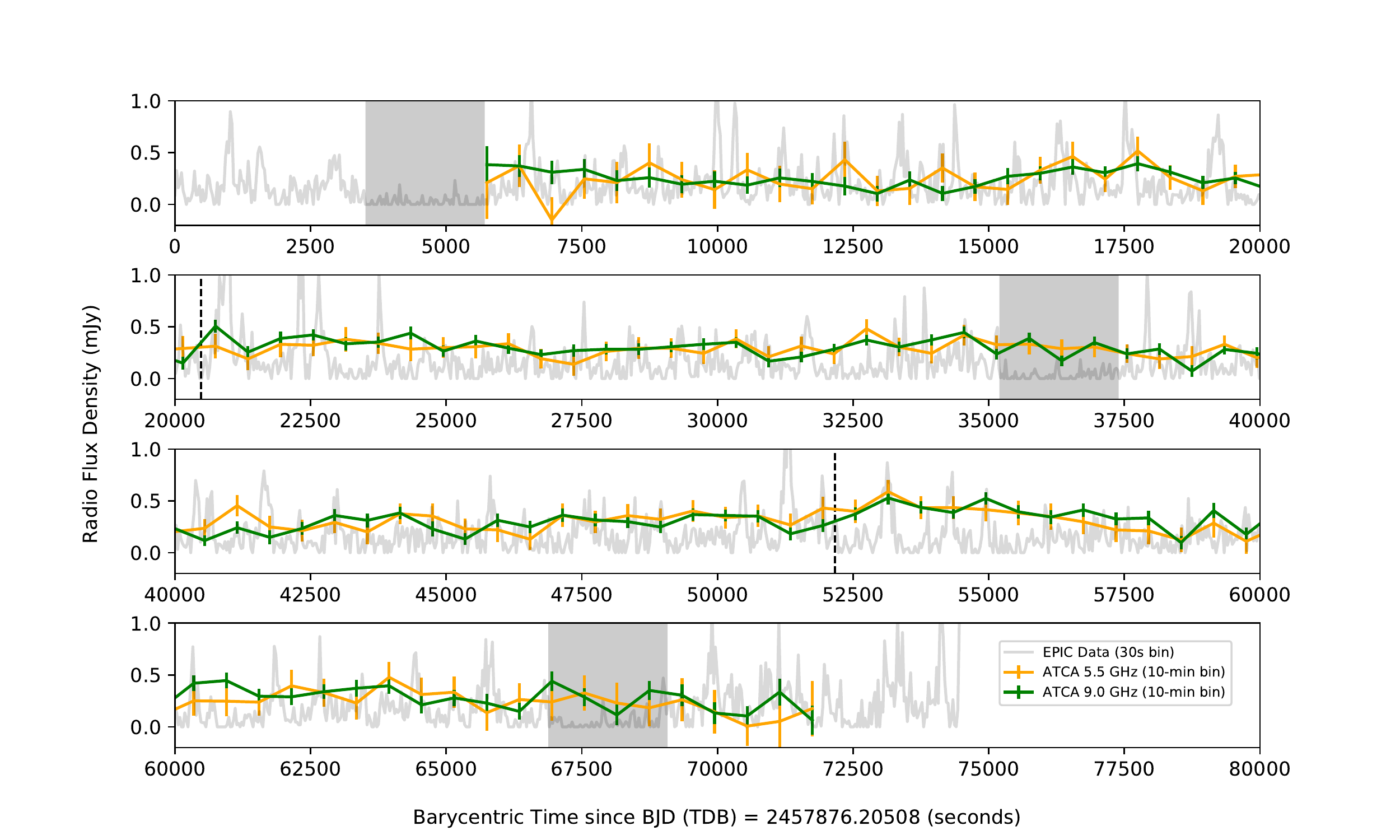}
\caption{A similar light curve as in Figure~\ref{fig:epic_lc}, but with the ATCA 5.5 and 9.0~GHz radio continuum data. The radio continuum data show modest variability but no clear flares, and there is no evidence that the radio emission is eclipsed.}
\label{fig:atca_lc}
\end{figure*}

While there was no significant 3--79~keV emission detected by \textit{NuSTAR} during the X-ray eclipses in 2016 (S16), we detect $\sim$50 net counts in the pn and MOS~1/2 combined image. Given the low source counts and the high background noise ($\sim$50 counts), we used \textit{W-statistic} (a modified version of \textit{C-statistic}; \citealt{1979ApJ...228..939C}), which is able to handle Poisson-distributed background spectra. 
We binned the spectra using \texttt{grppha} to at least one count per bin as suggested in the \texttt{XSPEC} manual. 
A simple absorbed power-law with the hydrogen column density fixed at the Galactic value of \nh\ $=3.66\times10^{20}$\cm\ was used as a phenomenological model. No intrinsic absorption is assumed because (i) the limited data quality does not allow a good probe for an intrinsic absorption, and (ii) the X-ray emission observed during the eclipse is likely already scattered away from the absorbing material in the disk.
In addition, no cross-calibration correction was applied on the spectral model (i.e., the cross-calibration factors were fixed to one) because of the low S/N. 
The best-fit photon index is $\Gamma=1.9^{+0.5}_{-0.4}$, with an inferred luminosity of $L_x\approx3\times10^{31}$\lum\ (0.2--10~keV), which is about 2\% of the X-ray luminosity in \textit{Average1}.

\begin{table*}
\renewcommand{\arraystretch}{1.5}
\scriptsize
\centering 
\caption{\texttt{pcfabs} fittings of \src\ (eclipse emission subtracted)}
\begin{tabular}{llrccccr}
\toprule
Dataset & Model & \multicolumn{1}{c}{\nh} & $C_2$ & Fraction & $\Gamma$ & $L_{\rm nth}$ & \multicolumn{1}{c}{$\chi_\nu^2$} \\
 & & \multicolumn{1}{c}{($10^{22}$\cm)} & & (\%) & & ($10^{33}$\lum) & \\
\hline

\textit{Low1} & \texttt{pcfabs*pow} & $11.9^{+0.8}_{-2.6}$ & $1.1^{+0.1}_{-0.1}$ & $100$ & $1.28^{+0.10}_{-0.11}$ & $0.99^{+0.53}_{-0.05}$ & 117.0/95 \\

\hline

\textit{Medium1} & \texttt{pcfabs*pow} & $9.6^{+0.8}_{-0.7}$ & $1.0^{+0.1}_{-0.1}$ & $98.3^{+0.9}_{-1.6}$ & $1.42^{+0.10}_{-0.10}$ & $3.86^{+0.22}_{-0.22}$ & 50.0/48 \\

\hline

\textit{High1} & \texttt{pcfabs*pow} & $4.8^{+0.6}_{-0.5}$ & $1.0^{+0.1}_{-0.1}$ & $96.6^{+1.7}_{-3.0}$ & $1.21^{+0.13}_{-0.14}$ & $5.63^{+0.36}_{-0.36}$ & 22.2/33 \\

\hline
\hline

\end{tabular}
\footnotetext{See the caption of Table \ref{tab:xmm_spec} for details.}
\label{tab:xmm_spec_eclipse}
\end{table*}

These relatively soft X-rays are possibly scattered from the atmosphere of the companion, or in an extended accretion disk corona (ADC; \citealt{1982ApJ...257..318W}). In the former case, the scattered emission is strong only during the eclipse when the pulsar is behind the scattering medium. The ADC emission, by contrast, is the weakest during the eclipse, and can be observed in all other phases. To examine this possible ADC component, we added the eclipse emission spectrum (fixed at the best-fit parameters) to the partially-absorbed power-law model, and fit the composite model to the flux-resolved spectra (Table \ref{tab:xmm_spec_eclipse}). 
In general, the fits are not improved. 
For the best-fit parameters, while there is almost no change on \textit{High1}, the power-law components of \textit{Low1} and \textit{Medium1} are significantly harder than the previous ones. In particular the \texttt{pcfabs} component of \textit{Low1} is no longer required (fraction = 1) as the eclipse emission cancels out the soft X-ray excess. However, we argue that this is likely a coincidence. Although an extended ADC is possibly observable during the eclipse, a large fraction of it is still occult. Much brighter ADC emission should therefore be observed in \textit{Low1} unless the ADC is extremely extended. For LMXBs, the ADC can extend up to $1 R_\sun$ in the radial direction \citep{2004MNRAS.348..955C}, which just slightly exceeds the companion of \src\ ($R_2=0.83\,R_\sun$; S16), not to mention that such an extended ADC requires a powerful central engine of $L_X\sim10^{38}$\lum\ \citep{2004MNRAS.348..955C} that is almost 1000 times higher than that of \src. 
Thus, the consistency between the soft X-ray excess in $Low1$ and the eclipse emission actually disfavors the ADC scattering scenario. 
As a result, we assumed the X-ray scattering off the companion's atmosphere as the origin of the eclipse emission, which would only bring a minor effect to our spectral analysis.

\section{Radio continuum Results}

Figure \ref{fig:atca_lc} shows the 5.5 and 9.0 GHz light curves of \src, overlaid on the simultaneous X-ray light curve for the same time period. As discussed above, the radio data are binned at time intervals of 10~min (600~sec). 

The main results from the radio data are as follows. First, \src\ is well-detected in both frequency bands at all times, with a mean out-of-eclipse flux density of $290\pm7$ and
$300\pm6 \mu$~Jy at 5.5 and 9.0 GHz, respectively. For a power-law spectrum with flux density $S_{\nu} \propto \nu^{\alpha}$, these values imply a mean $\alpha = 0.07\pm0.07$. Second, there is no evidence that the radio emission is eclipsed, with mean in-eclipse flux densities of $295\pm21$ and
$334\pm18 \mu$~Jy at 5.5 and 9.0 GHz. Hence the radio emission must primarily arise on size scales larger than the projected secondary ($\gtrsim 3.5 \times 10^{10}$ cm, using the parameters from S16). 

Next, while the radio continuum flux density is time-variable, no obvious flares can be discerned, quite unlike the X-ray and optical/UV light curves (Figure \ref{fig:atca_lc}). Given that the X-ray flares typically last for 10--40~sec and the radio light curve is binned on a time-scale of 600~sec, the lack of obvious flares in the radio light curves is perhaps not surprising. However, we do find some evidence that higher X-ray emission is associated with higher radio emission. If we consider the radio emission in the \textit{Low1}, \textit{Medium1}, and \textit{High1} X-ray categories described earlier, the \textit{Low1} and \textit{Medium1} categories have mean radio flux densities consistent with the full data set, while the \textit{High1} X-ray category is associated with a 5.5 GHz flux density of $427\pm36 \mu$Jy (brighter at $3.8\sigma$)
and a marginally steeper radio spectrum of $\alpha=-0.68\pm0.39$.

Table \ref{tab:atca} shows the flux densities as well as the spectral indices of these subsets of the radio data.

\subsection{Older radio data}

Here we briefly discuss the results from the Aug 2016 radio continuum observations of \src. These were not taken simultaneously with any X-ray observations.

We find a flux densities of $303\pm9$ and $337\pm8 \mu$Jy at 5.5 and 9.0 GHz, respectively, giving a spectral index of $\alpha = 0.21\pm0.08$. These values are entirely consistent with those measured in May 2017, and hence show that at least over the $\sim 9$ month separation of these epochs that the radio behavior of the binary is stable.

\begin{table*}
\centering 
\caption{ATCA radio properties of \src\ in May 2017}
\begin{tabular}{lllr}
\toprule
Dataset & 5.5~GHz & 9.0~GHz & Spectral Index \\
 & ($\mu$Jy) & ($\mu$Jy) & ($\alpha$)\\
\hline
In eclipse & $295\pm21$ & $334\pm18$ & $0.25\pm0.22$ \\
Out of eclipse & $290\pm7$ & $300\pm6$ & $0.07\pm0.07$ \\
\textit{Low1} & $281\pm8$ & $302\pm6$ & $0.15\pm0.10$ \\
\textit{Medium1} & $307\pm20$ & $268\pm16$ & $-0.28\pm0.22$ \\
\textit{High1} & $427\pm36$ & $305\pm27$ & $-0.68\pm0.39$ \\
\hline
\end{tabular}
\label{tab:atca}
\end{table*}

\section{Discussion}
Individual X-ray flaring events are not uncommon in redback and black widow systems, e.g., PSR~J1048+2339 \citep{2018ApJ...866...71C,2019A&A...621L...9Y}, 3FGL~J0838.8$-$2829 \citep{2017ApJ...844..150H}, and PSR~J1311$-$3430 \citep{2012ApJ...754L..25R,2015ApJ...804..115R,2017ApJ...850..100A}, but this paper is the first to show evidence for a system with a fully flare-dominated accretion mode. Perhaps the closest comparison is the few flare-dominated epochs of PSR~J1023+0038 in its sub-luminous disk state \citep{2014ApJ...791...77T,2014ApJ...797..111L,2015ApJ...806..148B,2019ApJ...882..104P}, though these have typically been short-lived, and make up only a small fraction of the observed modes in the current accretion state of PSR~J1023+0038. By contrast, \src\ maintained this flare-dominated state during the entirety of our $\sim70$~ksec observations in May 2017, and despite the lower sensitivity of earlier X-ray observations, appears to have been in a similar state in May 2016 and likely much earlier (see S16). 

\subsection{The concurrent optical/UV flares}
Given the large dispersion in the flare amplitudes between X-rays and optical/UV (Figure \ref{fig:sigsig}), the emission mechanisms of the two bands are probably different. 
Another intriguing property is that the optical/UV flares are generally longer than the X-ray flares (Figure \ref{fig:sigsig}), implying that that the optical/UV flares are emitted from the more outer region (e.g., the accretion disk). 

Perhaps the simplest model is that the optical/UV flares are the reprocessed emission from the X-rays, for example, due to ``reflection'' of the accretion disk. Considering the light-travel time, there should be a minimum time delay in the optical/UV light curve of at least $\sim 2.3$~sec to the outer disk.
Our analysis found an insignificant delay of $4.5\pm6.8$~sec, which is consistent with (but does not constrain) this expected delay. We also note that several optical/UV flares appear to have started earlier than their X-ray counterpart (see Figures \ref{fig:16flare}h as the most prominent case), which would not be consistent with this simple model.
Coordinated multi-wavelength observations by an X-ray timing instrument such as \textit{NICER} and ground-based telescopes capable of fast photometry could clarify this in the future. 

\subsection{Intrinsic X-ray variability or rapidly varying \nh?}
Our flux-resolved X-ray spectral analysis showed an enhancement in \nh\ when the X-ray flux gets faint. This is reminiscent of variable absorption, instead of an intrinsic flux change, resulting in the strong variability observed. 
The idea has also been used to explain the X-ray variability of the edge-on LMXB, 47~Tuc-X5, observed by \textit{Chandra} \citep{2003ApJ...588..452H,2016ApJ...831..184B}. 
The variable \nh\ could be due to a precessing accretion disk that varies the obscuring gas in the line-of-sight. Optical emission from the outer disk region would then have a better chance than the X-rays from the inner part to be seen through a cloud gap. This naturally explains the non-symmetric relation between the X-ray and optical flares as well as the shorter flaring time-scales in X-rays. 

The immediate objection is that the spectral fits find significant variations of the unabsorbed X-ray flux and the photon index. Although this might be attributed to imperfect correction for the absorption if each spectral dataset still contains too wide a range of fluxes, it is questionable whether the effect is sufficient to remove the variations. Additionally, the X-ray and optical/UV flares would be at least weakly correlated, if they both originate from obscuration. 
However, we find no correlation in Figure \ref{fig:sigsig}. 
The contradiction is better revealed by Figure \ref{fig:16flare}h and \ref{fig:16flare}i, in which the two optical flares appear very different in brightness and duration while the X-ray profiles are nearly identical. It is also unclear how the absorbing gas becomes so clumpy to form cloud gaps very frequently. 
Taken as a whole the variable absorption scenario is unable to explain the flaring state in many ways, and so we rule the possibility out. 

\subsection{Thermal scenarios for the soft X-ray excess}
\label{sec:ther}
As demonstrated in \S\ref{tab:xmm_spec}, the soft X-ray excess of the saddle-shaped spectra can be modelled by a single-temperature blackbody (a multi-temperature disk has been ruled out as the inferred inner radius is too small). 
If this thermal component is genuine, the neutron star surface will be the most reasonable origin of the emission. 
Though it is still unclear whether tMSPs are accretion-powered (e.g., \citealt{2015ApJ...806..148B,2015ApJ...807...33P}) or rotation-powered (e.g., \citealt{2014ApJ...785..131T,2016ApJ...830..122J,2017NatAs...1..854A,2019ApJ...882..104P}) during the sub-luminous disk state, such thermal emission is actually possible in both cases. 

For accretion-powered pulsars, hotspots can be formed on the neutron star surface when the magnetically channelled accretion flows heat up the magnetic poles. The temperatures of the hotspots are around 0.1--1~keV, which are consistent with the thermal component (blackbody or \texttt{nsa}) of \src. 
However, the apparent size of the blackbody (i.e., $\sim0.1$~km; Table \ref{tab:xmm_spec}) is much smaller than the typical size of the hotspots seen in AMXPs (i.e., radius of a few km; \citealt{2002MNRAS.331..141G,2005MNRAS.359.1261G}) and tMSPs (e.g., $\approx3$~km for PSR~J1023+0038 in the high mode; \citealt{2015ApJ...806..148B}). For the \texttt{nsa} fits, despite the larger emission sizes inferred, most of them are still less than 1~km. \textit{Medium1} is the only dataset that yields an emission region larger than 1~km, but the statistical uncertainties are also huge making the case marginal. 

For rotation-powered pulsars, X-ray emitting regions with temperatures of 0.1--1~keV can be created by polar cap heating (see \citealt{2002ApJ...568..862H} and the references therein). These heated polar cap regions are expected to be large ($\sim1$~km in radius for MSPs), which is, again, too big for the thermal component of \src. 
Alternatively, the thermal X-rays could be generated by the back-flow of the primary charged particles from the outer gap. The heated region would be much smaller in this scenario (i.e., $\sim0.1$~km in radius; \citealt{2003A&A...398..639Z}). However, such thermal components should be accompanied by another slightly cooler (i.e., $\lesssim0.1$~keV) but larger (i.e., $\sim1$~km) blackbody component, which is not seen in \src. 

Apart from the size inconsistency, the weak X-ray absorption for the thermal component is unexplainable. If the soft X-ray photons are really coming from the pulsar surface (the innermost observable region of the system), the thermal component should be highly absorbed. However, no intrinsic absorption is observed for the thermal emission, in contrast to the strong intrinsic absorption found for the non-thermal component. Besides, the inferred photon indices of the non-thermal component are significantly harder ($\Gamma\approx0.9$--1.1) than that of the two known tMSPs in the sub-luminous state ($\Gamma\approx1.6$--1.8; \citealt{2013A&A...550A..89D,2014ApJ...797..111L,2015ApJ...806..148B}), while the partially-absorbed power-law gives more reasonable results in this sense, especially in \textit{Low1} and \textit{Medium1}} ($\Gamma\approx1.8$ and 1.6, respectively; Table \ref{tab:xmm_spec}).

Based on the above arguments, we conclude that an extra thermal component as the origin for the soft X-ray excess is highly unlikely. 

\subsection{Non-thermal scenarios}
For a more physical picture, we considered the ``propeller'' scenario \citep{1975A&A....39..185I}, which has been widely used to understand the mode switching phenomenon as well as the high-energy emission observed in tMSPs \citep{2015ApJ...807...33P,2015ApJ...807...62A,2016A&A...594A..31C}. 
In the so-called propeller regime, where the accretion disk is truncated by the pulsar magnetosphere outside the co-rotation radius (i.e., $r_{\rm m}>r_{\rm c}$), most of the inflowing material is ejected by the centrifugal barrier, but a small fraction of the gas can still be accreted onto the neutron star through the magnetic field \citep{2015MNRAS.449.2803D}. 
Recent magnetohydrodynamics (MHD) simulations have shown that this partial accretion process can be possible for neutron star systems \citep{2014MNRAS.441...86L}. 
The X-ray pulsations of PSR~J1023+0038 and PSR~J1227$-$4853 detected in the high mode \citep{2015ApJ...807...62A,2015MNRAS.449L..26P} could be evidence for this partial accretion. 

As mentioned in \S\ref{sec:intro}, the low, high, and flare accretion modes are common in PSRs~J1023+0038 and J1227$-$4853. 
In Figure \ref{fig:epic_lc}, we compare the X-ray light curve of \src\ with the X-ray luminosities of PSR~J1023+0038 in the three modes \citep{2015ApJ...806..148B}. 
The ``quiescent'' state of \src\ has a similar luminosity to the low mode of PSR~J1023+0038. No obvious high mode of \src\ is seen in the \textit{XMM-Newton} light curve, but some weak X-ray flares are comparable to the high-mode flux of PSR~J1023+0038 (e.g., the fourth X-ray flare just after $t=2500$~sec in Figure \ref{fig:epic_lc}). 

One possibility is that these weak flares represent a \textit{transitory} high mode in \src. 
It has been suggested that the high and low modes are referring to a tMSP system staying in or away from the propeller regime, respectively \citep{2016A&A...594A..31C}. 
In this context, \src\ might stay in the propeller regime with a short duration of $\sim10$--$40$~sec in typical flaring episodes. 
This would also be in agreement with the likely small inner disk inferred in S16 (an inner radius of a few tens of kilometers, which is comparable to the co-rotation radius of a MSP). A relatively weak and/or unstable accretion flow of \src, which could not support a sustainable propeller state, would be a possible reason for the transitory high mode. 

There are numerous possible counterarguments to this picture: the concurrent X-ray and optical/UV variability of \src\ is quite unlike that for PSR~J1023+0038, where the optical flux has not been observed to follow the low/high mode switching \citep{2015ApJ...806..148B}. 
Optical variability that is analogous to the X-ray mode switching was found in the light curves of PSR~J1023+0038 though \citep{2015MNRAS.453.3461S,2018MNRAS.477..566S,2018MNRAS.474.3297H}.
In addition, a negative correlation is seen between the radio and X-ray luminosities in the low and high modes of PSR~J1023+0038 \citep{2018ApJ...856...54B}, but the correlation is likely positive in \src. The \nh\ variation could be another issue: strong absorption is generally expected if the X-ray emission originates from a more inner region in the high mode, but \src\ is being the opposite. Further, the partial X-ray absorption would be hard to explain, if the emission region is tiny (a few tens of kilometers)---the emission would likely be ``fully'' absorbed. 

Alternatively, the absence of the high mode in \src\ could also mean that the system did not enter in the propeller regime at all. 
Provided that the rotation-powered activity of the radio/$\gamma$-ray pulsar turned on to push the inner edge of the disk away from the light cylinder during the entire \textit{XMM-Newton} observation, the emission coming from the intrabinary shock between the relativistic pulsar wind and the accretion flow would take over in the X-ray band \citep{2014ApJ...785..131T,2014ApJ...797..111L,2016A&A...594A..31C}. 
Perturbation of the shock front due to the instability in the accretion flow can produce X-ray variability on a time-scale of $\sim100$~sec \citep{2014ApJ...785..131T}, which could be the origin of the flares. 
The concurrent optical/UV variability could then be attributed to the instability of the accretion disk that triggers the X-ray flares. 
Depending on the momentum ratio between the accretion flow and the pulsar wind, the intrabinary shock radius of PSR~J1023+0038 could be $\sim10^{10}$ -- $10^{11}$~cm during the sub-luminous disk state (a few thousands times larger than the co-rotation radius of a MSP; \citealt{2014ApJ...785..131T,2014ApJ...797..111L}).
If \src\ has a similar shock size, the partial X-ray absorption feature can be understood more easily. 
However, the large shock requires an inner disk radius of $\sim10^9$~cm \citep{2014ApJ...797..111L}, which is inconsistent with the result of S16 based on the optical light curve modelling. 
The problem regarding the \nh\ variation also stays unsolved in this theoretical frame. 

Models have also posited different explanations for the (less frequent) flares observed in other tMSPs. For example, in the tMSP model of \citet{2019ApJ...884..144V}, flares are caused by temporary increases in the cross-section of the pulsar wind/disk interaction. However, the origin of these variations is not obvious and does not give a straightforward prediction for the frequency or length of the flares observed in \src.

\begin{figure*}
\centering
\includegraphics[width=0.9\textwidth]{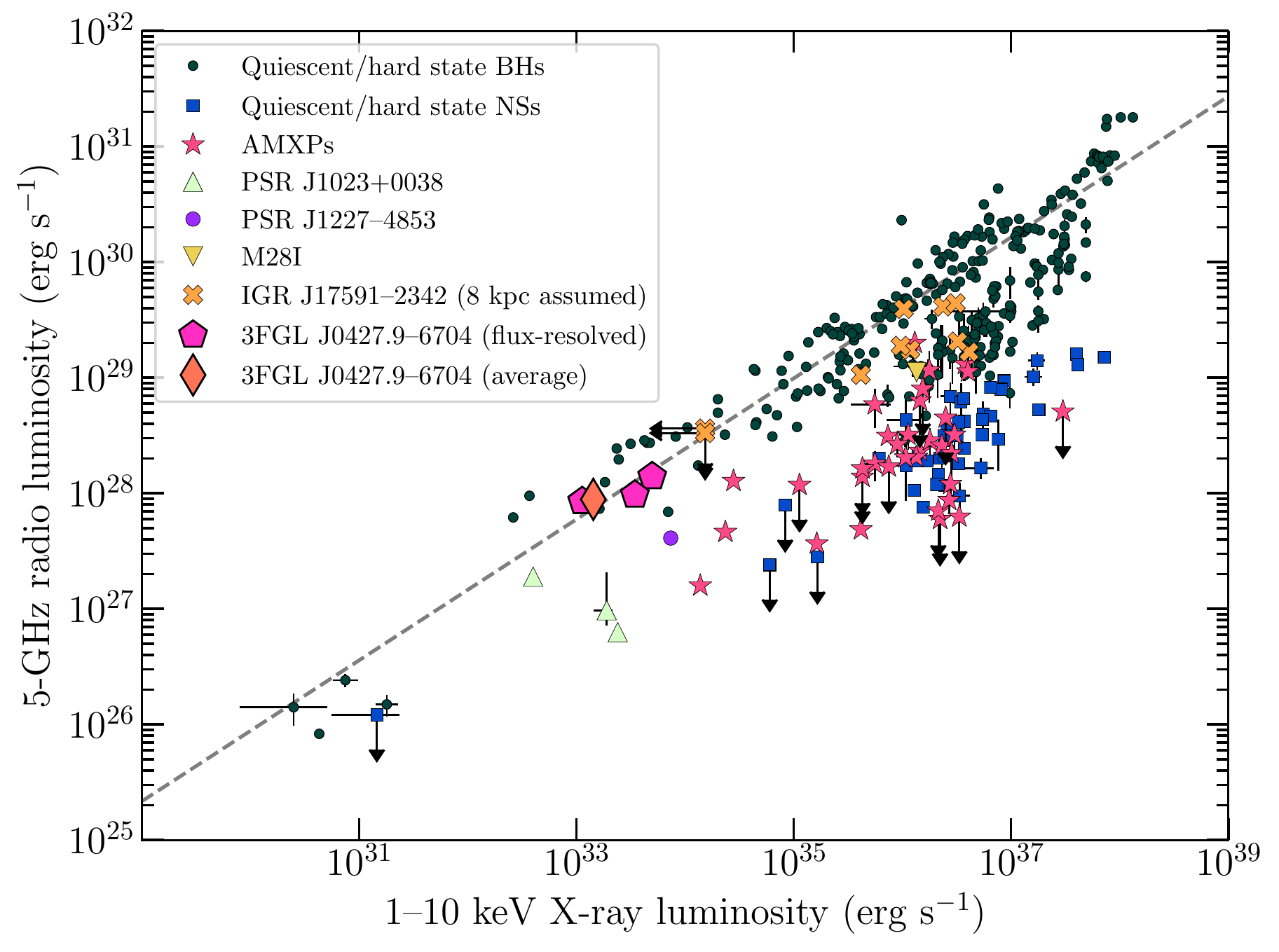}
\caption{A $L_X$--$L_R$ plot that contains \src, IGR J17591--2342 \citep{2020MNRAS.492.1091G}, three tMSPs, and other low-mass X-ray binaries obtained from the database of \cite{arash_bahramian_2018_1252036}.
\src\ sits close to the black hole relation at $L_X \sim 10^{33}$ erg s$^{-1}$ and is the only low-$L_X$ neutron star yet observed to do so.\\}
\label{fig:lrlx}
\end{figure*}

\subsection{Radio--X-ray correlation}

Accreting black holes show nearly ubiquitous radio continuum emission in their low/hard states, usually associated with a partially self-absorbed compact jet. However, 
our understanding of radio emission from accreting neutron stars has been slower to develop, with evidence emerging in the last decade of a much more complex situation for neutron stars than for black holes. It is clear that neutron stars do not follow a single relation between radio and X-ray luminosity, but show a wide range of radio loudness at
all $L_X > 10^{34}$ erg s$^{-1}$ (e.g., \citealt{2006MNRAS.366...79M,2011MNRAS.415.2407M,2017MNRAS.470..324T,2018MNRAS.478L.132G,2020MNRAS.492.1091G}).

The radio/X-ray correlation for neutron stars at a range of $L_X$ can help distinguish among physical models for the accretion flow. It also bears on the practice of using the
radio loudness of an accreting compact object to distinguish the nature of the accretor---typically neutron star vs.~black hole---at both low (e.g., \citealt{2012Natur.490...71S}) and high (e.g., \citealt{2019ApJ...883...39L}) X-ray luminosity.

Figure \ref{fig:lrlx} shows the radio/X-ray correlation for known black holes and neutron stars. Previous radio continuum studies of tMSPs have shown that these systems reliably show radio emission in the sub-luminous disk state \citep{2011MNRAS.415..235H,2013Natur.501..517P,2015ApJ...809...13D,2019PhDT.......J}. These few published tMSPs appear to sit on a track which is parallel to the black hole radio/X-ray correlation, but a factor of few fainter.

Remarkably, \src\ sits directly on the black hole correlation rather than the tMSP correlation. Since its radio continuum flux density depends only weakly on X-ray luminosity,
this statement is true both in the time-averaged sense and for subsets of the data selected by X-ray luminosity. The robustness of its location is unlike PSR~J1023+0038, where the mean radio/X-ray ratio is consistent with the proposed tMSP correlation, but the inverse behavior of radio and X-ray during mode switching means that the source is closer to the black hole correlation in the low mode, but closer to an extension of a ``hard state" neutron star correlation in the X-ray high mode (Figure \ref{fig:lrlx}).

The only other accreting neutron star shown to sit close to the black hole radio/X-ray correlation is the AMXP IGR J17591--2342, but this is at much higher $L_X \gtrsim 10^{35}$ erg s$^{-1}$, and its distance is also not yet well-constrained \citep{2020MNRAS.492.1091G}. This papers points out that there is no obvious reason why this AMXP should be much more radio-loud than other similar systems, and some candidate explanations such as the spin rate of the neutron star or possible beaming are not consistent with the data.

While the radio continuum emission in the sub-luminous disk state of PSR~J1023+0038 was initially mooted as arising in a jet \citep{2015ApJ...809...13D}, 
the extreme radio variability observed on short time-scales led \citet{2018ApJ...856...54B} to conclude that a steady jet could not be present. Instead, they suggest the radio emission could arise from expanding plasma bubbles at the interface of the pulsar magnetosphere and the inner disk.

By contrast, the radio emission from \src\ has properties more consistent with jets observed for black holes of a similar $L_X$: it is comparably radio-luminous, has a flat spectrum, is of substantial spatial extent ($>> 3.5 \times 10^{10}$ cm), and is relatively stable on time-scales of hours to months. While these properties do not prove that the radio emission arises from a jet rather than some other sort of sustained outflow, they are consistent with what one would expect for a jet. 

\subsection{Other flare-dominated sources?}

Here we briefly discuss another system that shares some properties with \src. In the globular cluster NGC~6652, the second-brightest source (here referred to as NGC 6652B)
has $L_X \sim 10^{34}$ erg s$^{-1}$ and shows flare-like variability on time-scales of a few $\times 100$~sec in a 2011 47~ksec \textit{Chandra} observation \citep{2012ApJ...751...62S}. Optical photometry of the source also displays rapid variability \citep{2012ApJ...747..119E}. While NGC 6652B is undoubtedly more luminous than \src, the phenomenology is sufficiently similar to be worthy of further study.

\section{Conclusion}
The observed properties of our simultaneous \textit{XMM-Newton} and ATCA observations of the edge-on tMSP candidate, \src\ are summarised as follows:

\begin{enumerate}
\item The X-ray variability seen in the 2016 \textit{NuSTAR} observation is resolved by \textit{XMM-Newton} EPIC. The variabilty is caused by vigorous X-ray flaring of the LMXB. While the flare mode accretion is occasionally seen in PSRs~J1023+0038 and J1227$-$4853, the accretion state of \src\ is entirely flare-dominated at a high flare occurrence rate of $\sim2\,$ks$^{-1}$. As the flares disappear during the three pulsar eclipses (Figure \ref{fig:epic_lc}), we conclude that the flares originate from the accreting neutron star. Except for the eclipses, the flares do not show any orbital dependence (Figure \ref{fig:phased_flare}). 

\item The flares are observed simultaneously in X-rays and optical/UV by \textit{XMM-Newton} EPIC and OM. 
Almost all of the X-ray flares have a corresponding optical/UV counterpart, but not every optical/UV flare has an X-ray partner. 
No significant time offset is seen between the X-ray flares and their optical/UV counterparts, with a formal cross-correlation offset of $4.5\pm6.8$~sec (Figure \ref{fig:cross}).
For those paired up, the flare durations in optical/UV are longer than those in X-rays, and the flare amplitudes are only weakly correlated (Figure \ref{fig:sigsig}).
Nevertheless, the optical/UV emission appears to lead the X-ray emission in a few cases (e.g., the flares shown in Figure \ref{fig:16flare}h, i, and p). 

\item The X-ray spectra (average or flux-resolved) of \src\ are saddle-like with a significant soft X-ray bump below 2~keV (Figure \ref{tab:xmm_spec}). This feature can be modelled either by an ordinary absorbed power-law with an additional thermal component or a partially-absorbed power-law (Table \ref{tab:xmm_spec}). We ruled out multi-component models because of the unphysically small thermal emission size inferred and/or the absence of intrinsic absorption for the thermal X-rays. In the flux-resolved analysis with a partially-absorbed power-law, \src\ becomes spectrally harder (i.e., lower $\Gamma$) with lower \nh\ at higher luminosities. There could be a high-energy exponential cut-off at $E_{\rm c}\approx30$~keV based on the joint \textit{XMM-Newton}-\textit{NuSTAR} spectrum, but this conclusive is tentative as the two X-ray datasets were non-simultaneous.

\item We detect weak X-ray emission during the three X-ray eclipses, possibly due to X-ray scattering off the companion's atmosphere. 

\item We consider the X-ray and optical properties of \src\ in the context of the commonly discussed propeller and intrabinary shock scenarios for the known tMSPs, but find that none of these models does a great job at explaining the observed phenomenology.

\item Mostly steady radio continuum emission is seen in all our ATCA observations, and we find that the radio flux density is positively correlated with the X-ray luminosity. Unlike for the optical and X-ray, we observe no radio eclipses. The stable, spatially extended, flat-spectrum radio emission has properties consistent with a jet.

\item \src\ sits precisely on the black hole radio/X-ray correlation, proving that even at $L_X \sim 10^{33}$ erg s$^{-1}$, some neutron stars can be as radio-bright as black holes. 

\end{enumerate}

\begin{acknowledgements}
We acknowledge useful discussions with K.~Sokolovsky. This work was supported by NASA grant 80NSSC18K0382, NSF grant AST-1714825, and a Packard Fellowship.
KLL is supported by the Ministry of Science and Technology of the Republic of China (Taiwan) through grant 108-2112-M-007-025-MY3. JCAM-J is the recipient of an Australian Research Council Future Fellowship, funded by the Australian government. COH is funded by NSERC Discovery Grant RGPIN-2016-04602. This work is based on observations obtained with \textit{XMM-Newton} an ESA science mission with instruments and contributions directly funded by ESA Member States and NASA. The Australia Telescope Compact Array is part of the Australia Telescope National Facility which is funded by the Australian Government for operation as a National Facility managed by CSIRO.

\end{acknowledgements}
\textit{Facilities}: \facility{XMM, ATCA}

\bibliography{j0427b}

\end{document}